Thèse d'Habilitation à Diriger des Recherches - Université Lyon 1

# Le système moteur au cœur de la Décision et de l'Action

Gerard Derosiere

**Composition du jury**

| | | | |
|---|---|---|---|
| Julie Duque | PU | UCLouvain | Présidente |
| Boris Burle | DR CNRS | Aix-Marseille Université | Rapporteur |
| David Robbe | DR INSERM | Aix-Marseille Université | Rapporteur |
| Thomas Michelet | MCF | Université de Bordeaux | Rapporteur |
| Stéphane Thobois | PUPH | Université Lyon 1 | Examinateur |
| Stéphane Perrey | PU | Université de Montpellier | Examinateur |

Soutenue le 11 mai 2023

# 1. Sommaire

## 2. Recherches passées

### 2.1. Synopsis

Mes travaux de recherche sont ancrés sur le plan théorique dans une vision parallèle de la Cognition, qui postule que les régions sensorimotrices sont impliquées – en parallèle à d'autres structures – dans des processus cognitifs historiquement principalement associés à des régions cérébrales d'ordre supérieur (*e.g.*, processus décisionnels associés au cortex préfrontal). Sur le plan méthodologique, les projets que nous avons menés ont impliqué l'utilisation de plusieurs techniques chez l'Homme (*e.g.*, la stimulation magnétique transcrânienne à impulsion unique, pairée et répétitive [sp-, pp- et rTMS] (e.g., [1–7]) ; l'électromyographie [EMG] (e.g., [8,9]) ; l'électroencéphalographie [EEG] (e.g., [10,11]); la spectroscopie dans le proche infrarouge [NIRS] (e.g., [1,12,13]), l'Imagerie par Résonance Magnétique [IRM] (e.g., [14]), la stimulation par interférence temporelle [TIS] (e.g., [15])) ainsi que diverses approches d'analyse des données (*e.g.*, analyses spatiotemporelles des potentiels évoqués moteurs [MEPs], de permutation de Monte-Carlo sur des données EEG, d'apprentissage machine sur des données NIRS).

J'ai effectué un doctorat en co-tutelle internationale (octobre 2011 - octobre 2014) associant l'Institut EuroMov de l'Université de Montpellier en France (Prof. Stéphane Perrey) et le département d'ingénierie électronique de la National University of Ireland Maynooth (Prof. Tomas Ward). Mes travaux de doctorat couvraient deux disciplines : les neurosciences et l'ingénierie. Dans le cadre de la partie neurosciences, j'ai montré, en utilisant une combinaison de techniques de spTMS, de NIRS et de stimulation neurale, que les baisses de performance survenant au cours de tâches d'attention soutenue sont associées à une augmentation de l'activité des régions frontopariétales (traditionnellement impliquées dans l'attention focalisée) mais aussi du cortex moteur primaire (M1). De manière importante, ce surengagement contribue à expliquer l'impulsivité motrice résultant des déficits d'attention. Une partie de ces résultats a été publiée dans Cerebral Cortex [16]. En parallèle à ce surengagement, une désactivation des structures sensorielles impliquées dans la tâche pourraient expliquer la baisse de sensibilité perceptive associée aux déficits d'attention [17]. Dans le cadre de la partie ingénierie, j'ai démontré la faisabilité de détecter les déficits attentionnels chez l'Homme, basés sur l'application d'algorithmes d'apprentissage machine sur des données NIRS [18–21].

J'ai ensuite obtenu un poste de chercheur postdoctoral en novembre 2014 au laboratoire Cognition et Actions (CoActions) de l'Institut des Neurosciences à Bruxelles (UCLouvain). Supervisés par la Prof. Julie Duque, mes travaux postdoctoraux ont montré le rôle du système moteur dans l'intégration de variables dites décisionnelles, notamment la récompense, l'évidence sensorielle et l'urgence. En octobre 2021, j'ai obtenu un poste de chercheur temporaire du Fonds National de la Recherche Scientifique (FNRS). Ce poste m'a permis de lancer une nouvelle ligne de recherche à l'Institut de Neuroscience de l'UCLouvain, sur la prise de décision basée sur l'effort et l'apathie. En janvier 2022, j'ai candidaté au concours INSERM de Chargé de Recherche et ai obtenu un poste au Centre de Recherche en Neuroscience de Lyon (CRNL). Depuis novembre 2022, j'établis ma ligne de recherche au CRNL sur la prise de décision basée sur l'effort et l'apathie. Les sections suivantes se concentrent sur une sélection d'études, que j'ai principalement accomplies au cours des huit dernières années, en tant que chercheur postdoctoral.

## 2.2. Cadre théorique

Comme mentionné plus haut, mes travaux se fondent sur une vision parallèle de la Cognition, que j'applique à mon objet d'étude principal : la prise de décision. Cette vision contraste avec une conception plus traditionnelle considérant les structures sensorielles et motrices comme majoritairement – si pas exclusivement – responsables des processus de Perception et d'exécution de l'Action, respectivement. En effet, les bases neurales de la prise de décision ont été historiquement considérées au travers d'un cadre théorique considérant le comportement comme résultant d'une progression sérielle de processus dits perceptif, cognitif et moteur. Dans cette optique, certaines structures comme le cortex orbitofrontal, le cortex préfrontal ventromédian, ou encore le cortex préfrontal dorsolatéral, implémenteraient dans un premier temps le processus cognitif – la prise de décision – avant de signaler aux structures du système moteur, comme M1, l'action à exécuter.

Cependant, au cours des années 2000, différentes données expérimentales ont remis en question la validité de cette vision sérielle, notamment dans le cadre de la prise de décision motrice, montrant par exemple que certaines structures du système moteur présentent également des changements d'activité neuronale pendant la phase de décision, alors qu'il n'y a pas de mouvement à exécuter [22,23]. Ce type de résultat a conduit à proposer que les structures motrices ne seraient pas uniquement impliquées dans l'exécution de mouvements mais spécifieraient continuellement les possibilités d'action offertes par l'environnement [24,25]. Selon cette vision alternative, la décision résulterait de processus neuraux survenant *en parallèle* au sein des structures sensorimotrices et des autres structures cérébrales. Dans cette optique, les possibilités d'action présentes dans l'environnement évoqueraient notamment une augmentation d'activité au sein de différentes populations de neurones du système moteur et l'engagement dans un mouvement donné émergerait lorsque l'activité aurait atteint un seuil de déclenchement dans l'une de ces populations. Ainsi, les changements d'activité observés au sein du système moteur joueraient un rôle décisif dans le processus de décision et détermineraient si une action est éventuellement sélectionnée et exécutée ou non. Une multitude de variables décisionnelles biaiserait ces changements d'activité motrice, rapprochant ou éloignant l'activité du seuil de déclenchement de l'action, et augmentant ou diminuant donc la probabilité d'initier certaines actions. Ces variables incluraient notamment la récompense associée aux différentes actions, l'évidence sensorielle en faveur de chacune des actions, et le niveau d'urgence dans un contexte donné. Comme décrit dans les sections suivantes, une grande partie de mes travaux a eu pour objectif de tester cette hypothèse.

## 2.3. Contribution de M1 à l'apprentissage par renforcement et à l'intégration de signaux de récompense pendant la prise de décision

Supervisions et publications

Ce projet a impliqué la supervision de deux étudiants de Master :

- Sophie Demaret, supervisée de janvier 2015 à juin 2016. Mémoire de Master sur l'implication causale de M1 dans l'implémentation de signaux de récompense pendant la prise de décision.
- Pierre Vassiliadis, supervisé de janvier 2015 à juin 2018. Mémoire de Master sur l'implication causale de M1 dans l'implémentation de signaux de récompense pendant l'apprentissage par renforcement et la prise de décision.

Articles liés au projet :

- **Derosiere G**, Zenon A, Alamia A, Duque J. (2017). Primary motor cortex contributes to the implementation of implicit value-based rules during motor decisions. *NeuroImage*, 146, 1115-1127 [26].
- **Derosiere G**, Vassiliadis P, Demaret S, Zenon A, Duque J. (2017). Learning stage-dependent effect of motor cortex disruption on value-based motor decisions. *NeuroImage*, 162, 173-185 [27].
- Vassiliadis P, **Derosiere G**. (2020). Selecting and executing actions for rewards. *The Journal of Neuroscience*, 40(34), 6474-6478 [28].

Rationnel et objectif

Les animaux de toutes espèces - y compris homo sapiens - présentent une caractéristique commune en termes de prise de décision : ils choisissent leurs actions afin de maximiser les bénéfices qui en résultent pour leur organisme. Dans le domaine des Neurosciences, ces bénéfices sont souvent désignés par le terme « récompenses ». En effet, le taux de récompense que reçoit un animal en milieu écologique au cours de sa vie joue un rôle fondamental dans sa longévité et sa fécondité [29], ce qui suggère qu'interagir avec l'environnement d'une manière qui maximise ce taux présente des avantages évolutifs directs. Pour cela, notre système nerveux a développé l'incroyable capacité d'apprendre, par un mécanisme associatif, quelles actions mènent aux récompenses les plus importantes dans l'environnement, un processus que l'on appelle « apprentissage par renforcement » [30,31]. Une fois la valeur de nos actions apprise, il devient possible, dans un environnement donné, de sélectionner les actions associées aux récompenses les plus élevées.

Au début de mon postdoctorat (en 2014-2015), plusieurs éléments semblaient suggérer que M1 pouvait directement contribuer à l'apprentissage de la valeur des actions et, subséquemment, à l'utilisation de cette information pendant la prise de décision. Premièrement, des études menées chez des primates non-humains et des rongeurs indiquaient que M1 reçoit des projections directes de structures dopaminergiques du mésencéphale impliquées dans l'apprentissage par renforcement, notamment de l'aire tegmentale ventrale et de la substantia nigra pars compacta [32–35]. Deuxièmement, des études de neuroimagerie avaient révélé que M1 présente des réponses neuronales phasiques suite à la présentation d'une récompense [36,37]. Ce type de réponse ressemblait fortement aux signaux d'erreur de prédiction de récompense (RPE), qui caractérisent les structures impliquées dans l'apprentissage par renforcement et la prise de décision. Enfin, certains chercheurs, qui avaient appliqué la spTMS sur M1 pour mesurer l'excitabilité corticospinale via les MEPs, avaient observé que l'amplitude de ces derniers augmente plus rapidement pendant la phase décisionnelle avant d'initier des actions associées à de fortes récompenses monétaires (*i.e.*, par rapport à des actions associées à de faibles récompenses ; [38,39]).

Cependant, ces résultats sur M1 restaient en marge des modèles existants sur le sujet qui impliquaient principalement des structures préfrontales, comme le cortex orbitofrontal [40–43], et sous-corticales, comme le striatum ventral [31,44–46], dans l'apprentissage par renforcement et dans l'intégration de signaux de récompense pendant la prise de décision. Surtout, ils ne permettaient pas de déterminer le rôle causal de M1 dans les processus d'apprentissage par renforcement et d'intégration des signaux de récompense pendant la décision. L'objectif de ce projet était donc de déterminer le rôle causal de M1 dans ces processus.

Approche méthodologique

Pour ce faire, avec deux étudiants de Master que je supervisais (Sophie Demaret et Pierre Vassiliadis), nous avons conduit deux études de rTMS, publiées dans *NeuroImage* [26,27]. Nous avons exploité un protocole de stimulation theta burst chez l'Homme sain nous permettant de moduler l'activité de M1 à différents stades de l'apprentissage par renforcement. Au sein des deux études, les sujets effectuaient une tâche de prise de décision dans laquelle ils devaient choisir un doigt donné (l'index ou le majeur de la main droite) en réponse à un stimulus donné (formes géométriques de différentes couleurs). La récompense monétaire obtenue après chacun des essais dépendait non seulement d'instructions explicites données aux sujets (*i.e.*, sélectionner le doigt adéquat face à une couleur ; *e.g.*, vert → index) mais aussi d'une règle implicite qui ne leur était pas divulguée (*i.e.*, sélectionner un doigt donné face à une forme précise ; *e.g.*, carré → index). En outre, les sujets obtenaient plus d'argent lorsqu'ils parvenaient à apprendre l'association implicite entre le mouvement d'un doigt et la forme des stimuli (*e.g.*, à sélectionner plus souvent l'index que le majeur en réponse au carré dans l'exemple ci-dessus). Notre but était de tester le rôle causal de M1 dans cet apprentissage.

Résultats et conclusion

Au sein de la première étude [26], les sujets (n = 56) ont effectué cette tâche pendant deux jours consécutifs. Nous avons appliqué la rTMS au milieu du deuxième jour, au moment où les sujets étaient sur le point d'intégrer la règle implicite (en plus des instructions explicites) dans leurs choix de réponses (*e.g.*, de sélectionner plus souvent l'index que le majeur en réponse au carré), comme cela était évident dans la proportion de choix du groupe contrôle, chez qui la rTMS était appliquée sur le cortex somatosensoriel droit. De manière intéressante, nos résultats montrent que les sujets ayant reçu une stimulation inhibitrice du M1 gauche (se projetant vers les doigts impliqués dans la tâche) n'intégraient pas la règle implicite dans leurs choix d'action en seconde moitié du jour 2. Par conséquent, ces derniers gagnaient moins d'argent que le groupe contrôle pendant la seconde moitié du jour 2 (Figure 1). A contrario, lorsque la stimulation avait pour effet de faciliter l'activité du M1 gauche, l'intégration de la règle implicite dans les choix d'action des sujets était améliorée et les sujets gagnaient plus d'argent. Ainsi, inhiber ou faciliter l'activité du M1 controlatérale aux effecteurs impliqués dans le choix d'action diminue et augmente, respectivement, la propension à sélectionner l'action la plus bénéfique en termes de récompense.

Au sein de la seconde étude [27], les sujets (n = 50) ont effectué cette tâche pendant trois jours consécutifs et nous avons appliqué une stimulation inhibitrice du M1 gauche à différents moments pendant ces trois jours, dont au milieu du jour 3. L'idée ici était de tester si l'implication de M1 variait au cours de l'apprentissage par renforcement et, spécifiquement, si elle perdurait une fois l'apprentissage de la règle implicite consolidé, en jour 3. De manière intéressante, nous avons observé que la stimulation n'avait plus d'impact significatif sur l'intégration de la règle implicite dans les choix des sujets en jour 3. Effectivement, les sujets ayant reçu une stimulation inhibitrice du M1 gauche en milieu de jour 3 présentaient un comportement décisionnel similaire au groupe contrôle, chez qui la rTMS était appliquée sur le cortex somatosensoriel droit en milieu de jour 3 également.

Dans l'ensemble, ces résultats indiquent que M1 intègre des signaux de récompense pendant la prise de décision. Cependant, sa contribution dans ce processus spécifique semble être plus importante lorsque la valeur des actions a été fraîchement acquise et tend à diminuer une fois l'apprentissage consolidé.

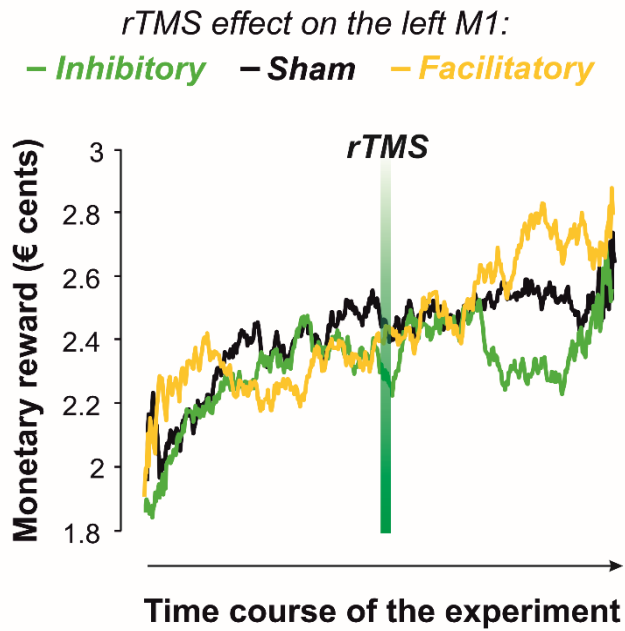

**Figure 1 : Inhiber ou faciliter l'activité du M1 controlatérale aux effecteurs impliqués dans le choix d'action diminue et augmente, respectivement, la propension à sélectionner l'action la plus bénéfique en termes de récompense** (figure adaptée de Derosiere et al., 2017b, *NeuroImage*) Au sein de cette première étude, la rTMS était appliquée en milieu du jour 2. Au sein des trois groupes de sujets, la récompense gagnée à chacun des essais (axe des ordonnées) augmentait jusqu'à la moitié du jour 2, dû à l'apprentissage de l'instruction explicite (*i.e.*, sélectionner le doigt adéquat face à une couleur). Au sein du groupe Sham (en noir), la récompense gagnée continuait d'augmenter ensuite, car les sujets du groupe Sham intégraient la règle implicite (en plus des instructions explicites) dans leurs choix de réponses (*e.g.*, de sélectionner plus souvent l'index que le majeur en réponse au carré). De manière intéressante, les sujets ayant reçu une stimulation inhibitrice du M1 gauche (en vert) gagnaient moins d'argent que le groupe contrôle pendant la seconde moitié du jour 2, car ils n'intégraient pas la règle implicite dans leurs choix d'action en seconde moitié du jour 2. A contrario, lorsque la stimulation avait pour effet de faciliter l'activité du M1 gauche (en jaune), les sujets gagnaient plus d'argent, indiquant que l'intégration de la règle implicite dans les choix d'action des sujets était améliorée.

## 2.4. Modulation de l'activité motrice par l'évidence sensorielle pendant la prise de décision

<u>Supervisions et publications</u>

Bien que non-officiellement impliqué dans la direction de son doctorat, ce projet a impliqué la supervision d'une partie du travail de doctorat de :

- <u>Andrea Alamia</u>, supervisé de mai 2017 à novembre 2018.

Articles liés au projet :

- **Derosiere G**, Klein PA, Nozaradan S, Zenon A, Mouraux A, Duque J. (2018). Visuomotor correlates of conflict expectation in the context of motor decisions. *The Journal of Neuroscience*, 38(44), 9486-9504 [10].
- Alamia A, Duque J, VanRullen R, Zenon A, **Derosiere G**. (2019). Implicit visual cues tune oscillatory motor activity during decision-making. *NeuroImage*, 186, 424-436 [47].

<u>Introduction du projet</u>

La vision est cruciale pour la survie d'une vaste majorité d'animaux diurnes, dont l'Homme. J'ai toujours été fasciné par la capacité de notre système nerveux à « filtrer », « amplifier » et « intégrer » certains signaux visuels afin de guider nos choix d'actions. En effet, face à l'abondance de stimuli présents dans notre environnement visuel, notre système nerveux a développé différents mécanismes permettant notamment *(1)* d'anticiper la survenue potentielle de conflit entre des stimuli multiples et *(2)* de traiter inconsciemment certains stimuli. Ces mécanismes lui permettent d'interagir de manière efficace avec l'environnement. Au sein de ce projet, nous avons conduit deux études EEG, publiées dans le *Journal of Neuroscience* [10] et dans

*NeuroImage* [11], afin d'étudier l'implémentation de ces mécanismes au sein des structures sensorimotrices corticales.

*2.4.1. Corrélats visuomoteurs de l'anticipation du conflit décisionnel*

<u>Rationnel et objectif</u>

L'abondance des stimuli dans notre environnement visuel génère souvent une difficulté à choisir l'action la plus appropriée, surtout lorsque des stimuli conflictuels appellent des actions incompatibles [48]. Imaginez, par exemple, un scénario de conduite dans lequel un feu de signalisation vient de passer au vert et un enfant traverse soudainement la route en courant. Dans de telles circonstances, l'action inappropriée (appuyer sur l'accélérateur) et l'action appropriée (appuyer sur la pédale de frein) entrent en conflit. Au niveau comportemental, la présence d'un tel conflit induit un coût, reflété par l'augmentation de la propension à sélectionner l'action inappropriée, et l'allongement du temps nécessaire à la sélection de cette action [49–52]. Au niveau niveau neuronal, le conflit produit une activation temporaire des représentations d'action inappropriée au sein du système moteur [53–58].

Dans certaines situations cependant, différents indices contextuels peuvent permettre de prédire l'apparition d'un conflit visuel [58,59]. Dans l'exemple précédent, un panneau d'école peut aider l'automobiliste à anticiper l'apparition soudaine d'enfants sur la route. Différentes études ont montré que, en augmentant l'anticipation du conflit, ces indices contextuels contribuent à diminuer la propension à sélectionner l'action inappropriée lorsque des stimuli conflictuels surviennent [56,58,60,61].

Lorsque je me suis intéressé à cette problématique au cours de mon postdoctorat (2016 - 2019), on pensait, au sein de la littérature, que cet effet comportemental résultait de l'amélioration d'une forme de contrôle « top-down » impliquant notamment le cortex frontal médian [62–65], qui présente une augmentation d'activité oscillatoire dans la bande thêta (4-8 Hz) lorsqu'un conflit est attendu [66]. Une hypothèse était que ce système de contrôle top-down permettait de réduire l'activation des représentations d'action inappropriée au sein des structures sensorimotrices pendant la prise de décision et donc, leur sélection. Cependant, très peu d'études avaient étudié l'impact potentiel de ce système de contrôle sur la compétition survenant au sein des structures sensorimotrices pendant la décision.

En réalité, une étude du laboratoire de la Prof. Julie Duque, dans lequel je travaillais, avait récemment montré que l'anticipation du conflit produit une suppression globale de l'excitabilité corticospinale pendant les décisions motrices [56,58]. Aussi, cette suppression globale semblait se produire de manière proactive puisqu'elle était déjà présente au moment de l'apparition du stimulus, avant même que les sujets n'aient perçu le stimulus visuel et sa possible nature conflictuelle. Par ailleurs, on savait très peu de choses sur l'impact de l'anticipation du conflit sur les structures sensorielles. L'objectif de notre étude était d'améliorer notre compréhension des changements sensorimoteurs sous-tendant l'anticipation du conflit visuel, en déterminant son impact sur la sélection des actions au sein du cortex moteur et son effet sur l'attention visuelle au sein du cortex visuel.

Approche méthodologique

Nous avons utilisé un EEG 64 électrodes chez des sujets humains (n = 20) pour identifier les corrélats visuels et moteurs de l'anticipation du conflit dans une version adaptée de la tâche d'Eriksen. Au sein de cette tâche, les sujets doivent choisir des mouvements de l'index droit ou de l'index gauche en fonction de la direction d'une flèche cible présentée au centre de l'écran. La flèche cible est entourée de distracteurs, représentés par des flèches pointant soit dans la même direction que la flèche cible (*i.e.*, dans les essais congruents) soit dans la direction opposée (*i.e.*, dans les essais incongruents ou conflictuels). Nous avons manipulé l'anticipation du conflit dans deux contextes en modulant la proportion d'essais incongruents par bloc d'essais. Dans les Mostly Incongruent Blocks (MIB), le conflit était présent dans 80 % des essais, produisant un contexte dans lequel les sujets pouvaient fortement l'anticiper, alors que dans les Mostly Congruent Blocks (MCB), le conflit n'était présent que dans 20 % des essais, ce qui diminuait son anticipation par les sujets.

En premier lieu, afin de quantifier l'impact de l'anticipation du conflit sur le système de contrôle « top-down » mentionné plus haut, nous avons analysé les modulations d'activité oscillatoire dans la bande thêta (4-8 Hz) au niveau des électrodes fronto-médiales dans chacun des contextes. Par ailleurs, afin de quantifier l'impact du contexte sur l'activité motrice, nous mesuré les potentiels évoqués bloqués en phase par rapport à l'exécution du mouvement, au niveau des électrodes centrales (nommés ci-après « response-locked potentials », ou RLP). Enfin, nous avons utilisé une approche permettant de quantifier les modulations d'attention visuo-spatiale en fonction du contexte (*i.e.*, MIB *vs.* MCB) et leur impact sur l'activité du cortex visuel. Pour cela, entre chaque essai, nous avons fait clignoter rapidement la localisation des différentes flèches sur l'écran, et ce, à différentes fréquences (*i.e.*, entre 11.5 et 16.5 Hz). Ce type de manipulation permet d'entraîner l'activité des neurones au sein du cortex visuel aux fréquences exploitées, évoquant des potentiels évoqués visuels stationnaires, ou steady-state visual evoked potentials (SSVEP), au sein du spectre fréquentiel des électrodes occipitales. L'amplitude du SSVEP à une fréquence donnée est modulée par l'attention dédiée à la localisation clignotant à cette fréquence. Ainsi, il nous était possible de quantifier les modulations d'attention visuo-spatiale en direction de la localisation de la cible et des distracteurs dans chacun des contextes.

Résultats et conclusion

En accord avec la littérature, nos résultats montrent que l'anticipation du conflit diminue la propension à sélectionner l'action inappropriée dans les essais incongruents. Cependant, cette amélioration se produit détriment de la vitesse de décision dans les essais congruents. Aussi, nous avons répliqué l'effet du contexte sur l'activité fronto-médiale (Van Driel et al., 2015), observant une augmentation d'activité dans la bande thêta (4-8 Hz) lorsque le conflit est attendu. De manière intéressante, nos résultats indiquent également que l'anticipation du conflit augmente l'activation des représentations des actions inappropriées au sein du système moteur pendant la phase de décision. En effet, l'amplitude des RLPs était plus grande au sein de l'hémisphère ipsilatéral à l'action exécutée dans les MIBs que dans les MCBs (Figure 2). Aussi, l'anticipation du conflit semble diminuer le filtrage des distracteurs au sein du cortex visuel ainsi que la focalisation de l'attention vers la localisation de la cible. En outre, les SSVEPs liés aux distracteurs étaient plus amples dans les MIBs que dans les MCBs (Figure 3), tandis que les SSVEPs liés à la cible étaient plus petits dans le premier type de bloc.

En conclusion, nos résultats montrent que l'anticipation du conflit ne réduit pas l'activation des représentations d'action inappropriée au sein des structures sensorimotrices pendant la prise de décision comme on le pensait au moment de la réalisation de cette étude. Au contraire, elle augmente la tolérance du système nerveux à présenter des activations sensorimotrices inappropriées sans que ces dernières ne mènent à sélectionner l'action inappropriée en question lorsque le conflit survient. Basé sur ces résultats, nous avons proposé un modèle alternatif postulant notamment que l'anticipation du conflit augmente la distance entre l'activité basale et le seuil de déclenchement de l'action au sein des structures sensorimotrices. Ce modèle permettrait d'expliquer pourquoi l'anticipation du conflit *(1)* diminue la propension à sélectionner l'action inappropriée dans les essais incongruents ; *(2)* allonge le temps de décision ; *(3)* génère une suppression globale de l'excitabilité corticospinale de manière proactive, avant même que les sujets n'aient perçu le stimulus visuel et sa possible nature conflictuelle (Klein et al., 2014 ; Duque et al., 2016) ; *(4)* augmente l'activation des représentations d'actions inappropriées au sein du système moteur pendant la phase de décision et *(5)* diminue le filtrage des distracteursau sein du cortex visuel.

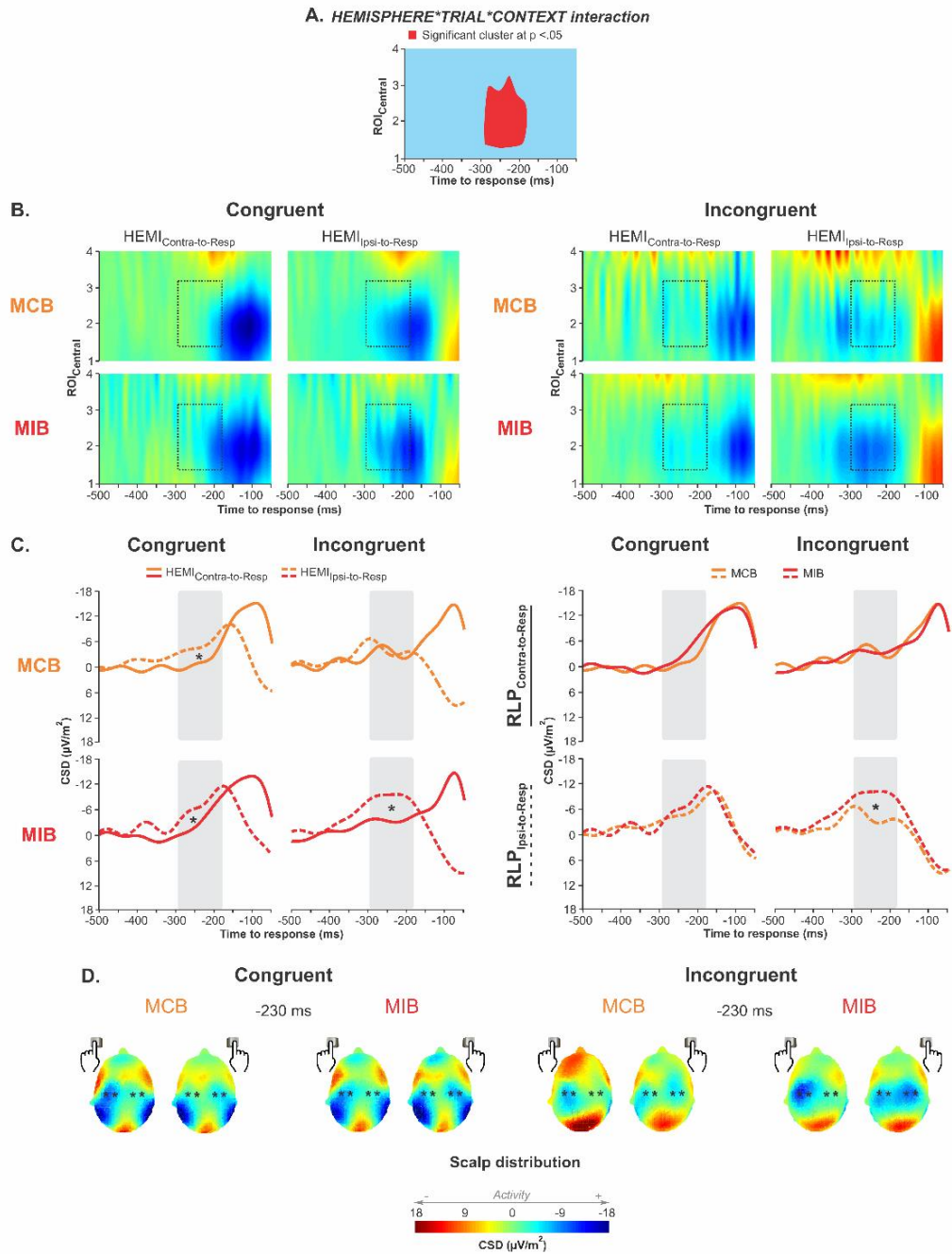

**Figure 2 : L'anticipation du conflit augmente l'activation des représentations des actions inappropriées au sein du système moteur pendant la phase de décision** (extrait de Derosiere et al., 2018, *The Journal of Neuroscience*). Le résultat principal peut être observé au sein de la colonne de droite de la partie C. Les tracés en pointillés représentent l'activité au sein de l'hémisphère ipsilatéral à l'effecteur choisi. Au sein des essais incongruents, cet hémisphère est celui qui est activé par

les distracteurs, appelant une action inappropriée. De manière intéressante, l'activation de cet hémisphère dans les essais incongruents pendant la prise de décision est supérieur dans le bloc MIC (en rouge) relativement au bloc MCB (en orange). Les autres parties de la figure représentent ce même résultat de différentes façons.

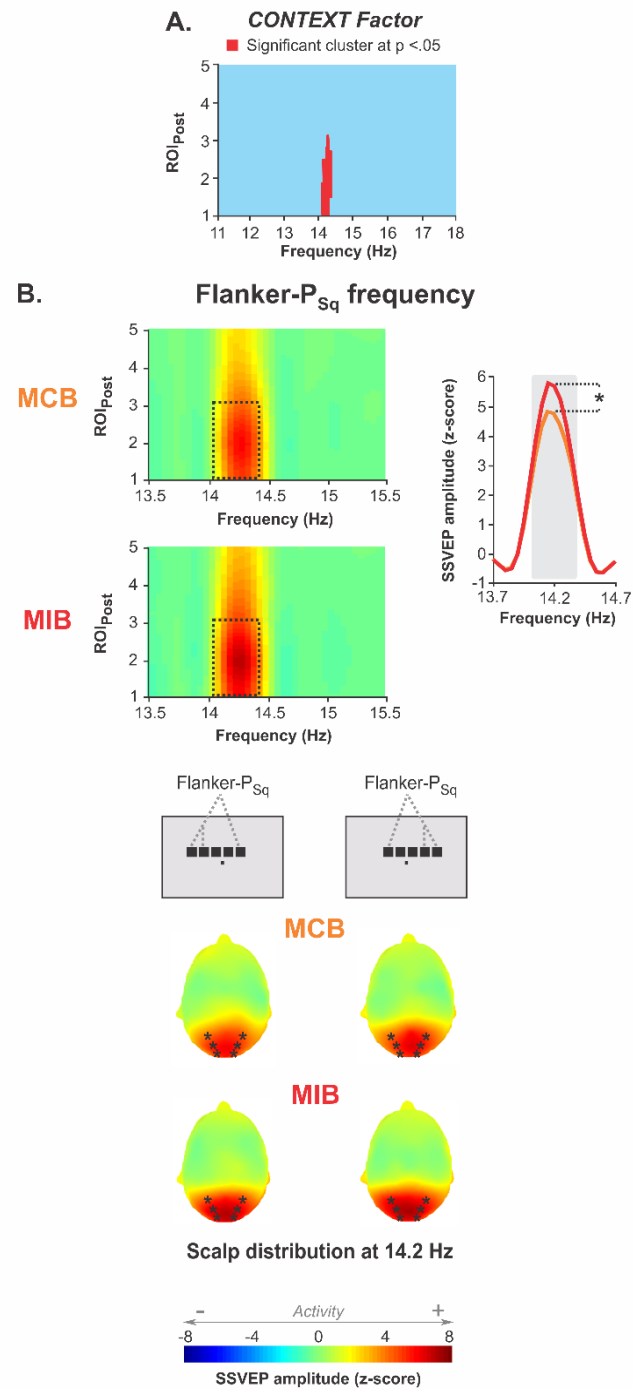

**Figure 3 : L'anticipation du conflit diminue le filtrage des distracteurs au sein du cortex visuel** (extrait de Derosiere et al., 2018, *The Journal of Neuroscience*). Les distracteurs (appelés Flanker-P ici), oscillaient à une fréquence de 14.2 Hz dans notre tâche. Les tracés sur la partie droite en B nous montrent que le spectre de fréquence

obtenu au niveau de la région d'intérêt (ROI) occipitale (*i.e.*, au niveau du cortex visuel) est plus ample dans les bloc MIB (en rouge) relativement aux bloc MCB (en orange), indiquant un filtrage attentionnel moindre dans le premier type de bloc par rapport au second.

### *2.4.2. Les informations visuelles inconscientes biaisent l'activité motrice pendant la prise de décision*

Rationnel et objectif

Pendant la prise de décision motrice, le M1 controlatéral à l'effecteur choisi présente une accumulation d'activité neuronale, reflétée notamment par une augmentation progressive de la désynchronisation des oscillations de fréquence beta (BFO, [15-35 Hz], souvent appelée event-related desynchronization [β-ERD]) [61,67–70]. La vitesse de cette accumulation dépend de la quantité d'informations – appelée « évidence sensorielle » en Neurosciences de la Décision – pointant en faveur de chacune des possibilités d'action. En effet, lorsque l'évidence sensorielle en faveur d'une action est forte, l'activité dans le cortex moteur controlatéral croît plus rapidement que lorsqu'elle est faible, augmentant ainsi la probabilité d'atteindre le seuil de déclenchement de cette action et donc de la choisir [71–75]. Cette modulation de l'activité motrice permet de comprendre pourquoi la vitesse et la précision des décisions sont toutes deux accrues lorsque l'évidence sensorielle favorise clairement une action [76].

Le système nerveux présente une capacité de traitement des stimuli visuels limitée, ne lui permettant pas de traiter tous les stimuli présents dans l'environnement avec le même degré de conscience. Par conséquent, une partie de l'information visuelle utilisée pour guider les décisions motrices reste implicite et n'accède pas à la conscience [77]. Alors que plusieurs études avaient démontré que l'évidence sensorielle influence l'accumulation d'activité motrice lorsque l'information est fournie par des stimuli explicites, perçus consciemment (Tosoni et al., 2008 ; Donner et al., 2009 ; Gould et al., 2012 ; Wyart et al., 2012), on connaissait mal les mécanismes neuronaux qui sous-tendent l'intégration de stimuli implicites, perçus inconsciemment, au sein du système moteur. L'objectif de notre étude était de mieux caractériser ces mécanismes neuronaux.

Approche méthodologique

Nous avons utilisé un EEG 32 électrodes chez des sujets humains (n = 18) pour quantifier l'impact de stimuli implicites sur l'activité oscillatoire du cortex moteur pendant la prise de décision. Les sujets ont effectué une tâche de random dot motion dans laquelle ils devaient choisir entre des mouvements de l'index droit et de l'index gauche en fonction de la direction dominante d'un nuage de points (droite ou gauche, respectivement). Aussi, le nuage de points pouvait être présenté dans trois couleurs différentes (*i.e.*, une couleur par essai). De manière importante, sans que les sujets n'en soient avertis, la couleur pouvait apporter de l'évidence sensorielle en faveur d'un des deux choix d'action (*i.e.*, mouvements de l'index droit ou gauche). En effet, deux des trois couleurs étaient fréquemment associées à une direction du nuage de points (dans 80 % des essais), apportant aux sujets de l'évidence sensorielle en faveur de la réponse correcte. De ce fait, lorsque ces couleurs étaient présentées dans les essais impliquant la direction opposée du nuage de points (*i.e.*, dans 20 % des essais), elles apportaient aux sujets de l'évidence sensorielle en défaveur de la réponse correcte, les incitant à choisir la réponse incorrecte. Des tests dédiés, réalisés en fin d'expérience, ont démontré que les sujets n'étaient pas conscients que la couleur était informative quant au choix d'action à apporter. Enfin, la troisième couleur était présentée dans proportions égales pour les deux directions de nuage de points et n'était pas informative quant à la réponse correcte. Nous avons réalisé des analyses temps-fréquence sur les signaux enregistrés au niveau des électrodes centrales afin de déterminer l'impact de l'évidence implicite sur l'activité oscillatoire du cortex moteur pendant la prise de décision.

Résultats et conclusion

Au niveau comportemental, nos résultats montrent que les sujets sélectionnent plus souvent l'action correcte et ce, avec un temps de réaction réduit, lorsque l'évidence sensorielle implicite favorise cette réponse. Par ailleurs, les sujets sélectionnent moins souvent l'action correcte lorsque l'évidence sensorielle implicite les incite à choisir la réponse incorrecte. En d'autres termes, bien que les sujets soient inconscients de l'information apportée par la couleur du stimulus (*i.e.,* que l'évidence sensorielle soit implicite), cette dernière influence directement leur comportement décisionnel.

Au niveau neurophysiologique, nos résultats montrent que l'évidence sensorielle implicite influence l'activité oscillatoire dans la bande de fréquence beta basse (16-25 Hz). En effet, la désynchronisation dans cette bande de fréquence est amplifiée au sein de l'hémisphère controlatérale à l'effecteur sélectionné lorsque l'évidence implicite favorise la sélection de l'action correcte (Figure 4). A contrario, la désynchronisation dans cette bande de fréquence est diminuée au sein de l'hémisphère controlatérale à l'effecteur sélectionné lorsque l'évidence implicite défavorise la sélection de l'action correcte. De manière intéressante, cette modulation de l'activité oscillatoire par l'évidence implicite est corrélée à la modulation de la vitesse de décision : au plus la désynchronisation dans la bande beta est amplifiée lorsque l'évidence favorise la sélection de l'action correcte, au plus les sujets réduisent leur temps de décision dans ces essais.

En conclusion, nos résultats apportent une extension des études antérieures, en indiquant que l'intégration d'évidence sensorielle d'origine explicite et implicite module l'activité motrice oscillatoire pendant la prise de décision. Ils démontrent donc un substrat neuronal – au sein du système moteur – fournissant un début d'explication à la manière dont certaines sources d'information inconscientes influencent les décisions humaines.

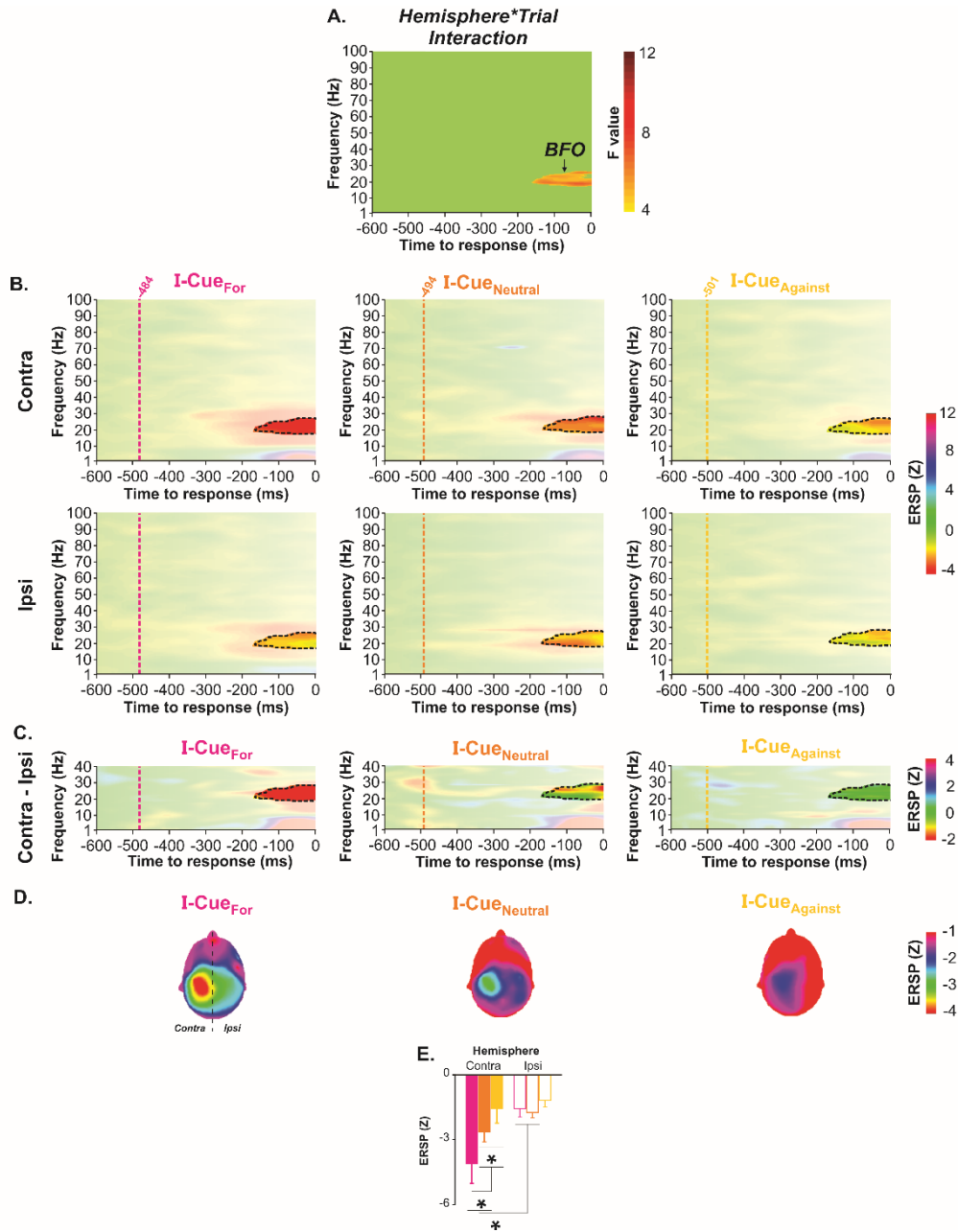

**Figure 4 : L'évidence sensorielle implicite influence l'activité oscillatoire de M1 dans la bande de fréquence beta basse (16-25 Hz)** (extrait de Alamia, *[…]* Derosiere, 2018, *NeuroImage*). L'effet le plus intéressant est synthétisé en partie E. Ici, la puissance spectrale mesurée dans la bande de fréquence béta basse (axe des ordonnées) a été mesurée au sein des hémisphères controlatéral et ipsilatéral à l'effecteur choisi (axe des abscisses ; barres colorées et blanches, respectivement). Les valeurs obtenues sont négatives, indiquant une diminution de puissance spectrale pendant la phase de décision par rapport à une ligne de base mesurée entre les essais – *i.e.*, une désynchronisation. De manière intéressante, cette désynchronisation est amplifiée au sein de l'hémisphère controlatéral à l'effecteur choisi lorsque l'évidence implicite favorise la sélection de l'action correcte (condition I-Cue$_{For}$, en magenta), relativement à la condition contrôle (I-Cue$_{Neutral}$, en orange). A contrario, la désynchronisation est réduite au sein de l'hémisphère controlatéral à l'effecteur choisi lorsque l'évidence implicite défavorise la sélection de l'action correcte (condition I-Cue$_{Against}$, en jaune), relativement à la condition contrôle.

## 2.5. Implémentation de signaux d'urgence et régulation du compromis vitesse-précision (speed-accuracy tradeoff ou SAT)

Supervisions et publications

Ce projet a impliqué la supervision de six étudiants de Master :

- Sara Lo Presti, supervisée de septembre 2018 à janvier 2019. Mémoire de Master sur la modulation de l'activité des représentations motrices de la jambe pendant des choix d'actions du membre supérieur.
- Roxanne Weverbergh, supervisée de janvier 2018 à juin 2019. Mémoire de Master en binôme avec C. Hermand (ci-dessous) sur la modulation de l'activité des représentations motrices de la main pendant des choix d'actions du membre supérieur.
- Caroline Hermand, supervisée de janvier 2018 à juin 2019. Mémoire de Master en binôme avec R. Weverbergh.
- Simon Vanhemelrijck, supervisé de janvier 2016 à juin 2017. Mémoire de Master les modulations d'activité de M1 pendant la régulation du SAT.
- Tristan Bouley, supervisé de janvier 2016 à juin 2017. Mémoire de Master en binôme avec A. Pepers sur l'implication causale de M1 dans la régulation du SAT.
- Antoine Pepers, supervisé de janvier 2016 à juin 2017. Mémoire de Master en binôme avec T. Bouley.

Articles liés au projet :

- **Derosiere G**, Thura D, Cisek P, Duque J. (2019). Motor cortex disruption delays motor processes but not deliberation about action choices. *Journal of Neurophysiology*, 122(4), 1566-1577 [78].
- **Derosiere G**, Thura D, Cisek P, Duque J. (2021). Trading accuracy for speed over the course of a decision. *Journal of Neurophysiology*, 126(2), 361-372 [79].

- **Derosiere G**, Thura D, Cisek P, Duque J. (2022). Hasty sensorimotor decisions rely on an overlap of broad and selective changes in motor activity. *PLOS Biology*, 20(4), e3001598 [80].
- Fievez F, **Derosiere G**, Verbruggen F, Duque J. (2022). Post-error slowing reflects the joint impact of adaptive and maladaptive processes during decision-making. *Frontiers in Neuroscience* [81].

Introduction du projet

La prise de décision est caractérisée par une covariation inhérente entre la vitesse et la précision des choix réalisés (*i.e.*, la propension à sélectionner une action correcte), faisant du compromis vitesse-précision (speed-accuracy tradeoff ou SAT) une propriété universelle du comportement animal. On ne peut y échapper. Néanmoins, nous présentons l'incroyable capacité de réguler volontairement notre SAT dans différents contextes, favorisant des stratégies de décision dites hâtives (*i.e.*, avec une vitesse de décision rapide, mais une propension réduite à sélectionner l'action correcte), ou prudentes (*i.e.*, avec une vitesse lente, mais une propension plus grande à sélectionner l'action correcte).

Il y a environ 10 ans, certains auteurs ont proposé des modèles computationnels postulant que la régulation du SAT est permise par l'ajustement d'un signal d'urgence [82]. Ce type de modèle assume que, au cours d'une décision, le système nerveux combine *(1)* l'évidence sensorielle perçue avec *(2)* un signal d'urgence, pouvant être modélisé comme une fonction linéaire augmentant progressivement au fil du temps. Cette augmentation linéaire de l'urgence diminue mécaniquement le niveau d'évidence sensorielle nécessaire pour atteindre le seuil de déclenchement de l'action. Au sein de ces modèles, un signal d'urgence avec un niveau initial élevé (*i.e.*, l'ordonnée à l'origine de la fonction linéaire) et une pente importante est associé à une vitesse de décision rapide et une propension réduite à sélectionner l'action correcte par rapport à lorsque le niveau initial et la pente de l'urgence sont moindres.

Une partie de mes travaux a eu pour objectif de mieux comprendre comment le système nerveux implémente ces signaux d'urgence afin de réguler le SAT. Ce travail a été réalisé en collaboration avec le Prof. Paul Cisek (Université de Montréal) et le Dr. David Thura (CRCN INSERM, Lyon) et a impliqué un séjour de deux mois en tant que chercheur visiteur à l'Université de Montréal en 2015. Dans une première étude exploitant la modélisation computationnelle, j'ai montré que la régulation du SAT pouvait se faire non seulement d'un contexte à un autre et d'une décision à une autre, mais aussi sur une échelle temporelle beaucoup plus courte, au cours même d'une décision. Alors que les études passées avaient modélisé l'urgence comme un signal linéaire, la modélisation du comportement des sujets montre que cette capacité de régulation rapide du SAT repose sur un signal d'urgence non-linéaire. Dans une seconde étude, j'ai combiné la rTMS et la modélisation computationnelle et ai montré que M1 contribue causalement à l'invigoration des processus moteurs survenant après le processus décisionnel mais pas à l'implémentation de signaux d'urgence pendant la période de décision, ce qui suggère que d'autres circuits cérébraux contrôlent ce dernier processus. Ces deux premières études ont été publiées dans le *Journal of Neurophysiology* [78,79]. La troisième étude est ma favorite pour plusieurs raisons conceptuelles et méthodologiques, et est celle sur laquelle je focaliserai la suite de cette section. Dans cette dernière étude, j'ai combiné différentes approches de spTMS et d'analyses de MEP pour cartographier dans M1 l'étendue des modulations d'activité pendant la régulation du SAT. Cette étude a été publiée dans *PLOS Biology*.

## 2.6. Les décisions sensorimotrices hâtives reposent sur une combinaison de modulations globales et spécifiques de l'activité de M1

Rationnel et objectif

Certaines études suggèrent que l'activité de M1 est amplifiée lorsque le contexte incite à prendre des décisions hâtives (*i.e.*, lorsque l'urgence est élevée) [70,83]. Aussi, la régulation du SAT en fonction du contexte implique des structures sous-corticales, en particulier les ganglions de la base [76,84] et le système noradrénergique [70,85], qui sont connues pour moduler l'activité de M1 de manière non-spécifique. Basé sur ces deux résultats, il a été proposé que la régulation du SAT repose sur une modulation non-sélective du gain neural au sein du système moteur, produites par les structures sous-corticales [70]. Ces structures amplifieraient globalement l'activité de M1 lorsque le contexte incite à prendre des décisions hâtives, indépendamment de la population de neurones finalement recrutée pour exécuter l'action au sein de M1. Cette amplification permettrait d'agir plus rapidement, aux dépens de la précision de décision.

Le pouvoir explicatif de l'hypothèse de modulation globale du gain neural a contribué à sa diffusion dans le domaine des Neurosciences de la Décision. Cependant, aucune donnée ne permettait de la confirmer ou l'infirmer directement. L'objectif de notre étude était donc de tester cette hypothèse.

Approche méthodologique

Cinquante sujets ont effectué une adaptation de la tâche des jetons (Cisek et al., 2009), dans laquelle ils devaient choisir entre des mouvements de l'index droit et gauche en fonction d'évidence sensorielle qui évoluait au cours du temps. Dans cette version de la tâche, les choix corrects menaient à une récompense monétaire qui était inversement proportionnelle au temps de décision, accentuant l'urgence d'agir (de +14 à +1 centime(s) d'euro). Surtout, les choix incorrects entraînaient une pénalité élevée (-14 centimes d'euro) ou faible (-4 centimes d'euro) dans deux contextes SAT différents (*i.e.*, deux types de bloc d'essais différents), incitant les sujets à privilégier des stratégies de décision prudentes ou hâtives, respectivement.

Afin de cartographier les modulations d'activité au sein de M1 pendant la décision, nous avons utilisé la spTMS et avons mesuré l'amplitude de MEPs dans 9 muscles des doigts et de la jambe. L'amplitude des MEPs était utilisée comme une mesure des variations d'excitabilité au sein des représentations motrices se projetant vers chacun des muscles. Plus spécifiquement, nous avons mesuré l'excitabilité de la représentation de l'index droit et gauche, qui était impliquée dans la prise de décision, du pouce et du petit doigt droits et gauches, qui n'étaient pas impliquées dans la décision et se trouvaient proches de celle de l'index en termes de somatotopie, et de trois muscles de la jambe droite, qui n'étaient pas impliquées dans la décision et se trouvaient loin de celle de l'index en termes de somatotopie.

Sur la base de ces mesures de MEP, j'ai développé deux nouvelles analyses. La première analyse permettait de visualiser sous forme de carte spatio-temporelle les variations d'amplitudes des MEPs au cours de la décision. Pour cela, j'ai organisé les données en fonction de la somatotopie du système moteur et du timing auquel les MEPs étaient mesurés pendant la décision (voir Figure 5, ci-dessous). La seconde analyse examinait la relation entre les changements d'excitabilité se produisant au sein de la représentation de l'index choisi et ceux qui se produisaient dans les autres représentations des doigts. Dans cette analyse, nous nous sommes concentrés sur les données obtenues au sein des représentations des doigts car elles étaient enregistrées simultanément dans chaque essai (*i.e.*, alors que les mesures au sein des représentations des jambes nécessitaient l'utilisation d'un autre type de bobine placée au niveau de la ligne médiane du crâne, à distance des représentations des doigts). Plus spécifiquement, nous avons cherché à savoir dans quelle mesure la variation d'amplitude des MEPs d'un essai à l'autre dans l'index choisi était liée à la variation d'amplitude des MEPs d'un essai à l'autre dans les autres muscles des doigts. Le raisonnement était le suivant : une corrélation positive élevée entre l'index choisi et des autres muscles des doigts indiquerait l'influence d'une source neurale commune sur leurs représentations, modulant les MEPs « en bloc ». En revanche, une corrélation faible, voire négative, indiquerait l'influence de sources affectant la représentation de l'index choisi de manière plus sélective et différenciée [86–90]. Nous avons comparé les valeurs de corrélation obtenues pendant la décision dans les contextes hâtif et prudent.

Résultats et conclusion

Au niveau comportemental, les sujets étaient plus rapides et moins précis quand la pénalité apportée pour des choix incorrects était basse par rapport à lorsqu'elle était élevée, indiquant qu'ils favorisaient une stratégie de décision hâtive. L'analyse computationnelle de ces données comportementales indique que cette régulation du SAT implique une augmentation de l'ordonnée à l'origine de la fonction d'urgence, diminuant le seuil de déclenchement de l'action dans le contexte hâtif.

Au niveau neurophysiologique, dans l'ensemble, nos données soutiennent l'idée la stratégie de décision hâtive est associée à une amplification de l'excitabilité de M1 non-sélective (Figure 5). En effet, dans le contexte hâtif, l'excitabilité motrice est amplifiée non seulement au sein de la représentation de l'index (impliquée dans la prise de décision ici) mais aussi au sein des représentations des muscles de la jambe. Toutefois, cet effet n'est pas global puisqu'il est limité au côté choisi et ne s'étend pas aux représentations du côté non choisi du corps. De manière intéressante, en plus de cet effet, nous avons également identifié une suppression locale de l'excitabilité motrice, entourant la représentation de l'index, également du côté choisi.

En conclusion, une stratégie de décision favorisant la vitesse par rapport à la précision semble impliquer différents types de modulation, produisant une amplification large (mais pas globale) et une suppression de l'excitabilité autour de la représentation exécutant l'action. Ce dernier effet peut contribuer à augmenter le rapport signal-sur-bruit de la représentation choisie, comme le suggère également les analyses de corrélation indiquant une plus forte différenciation des changements d'excitabilité entre la représentation de l'index choisi et les autres représentations motrices dans le contexte hâtif par rapport au contexte de prudence (Figure 6).

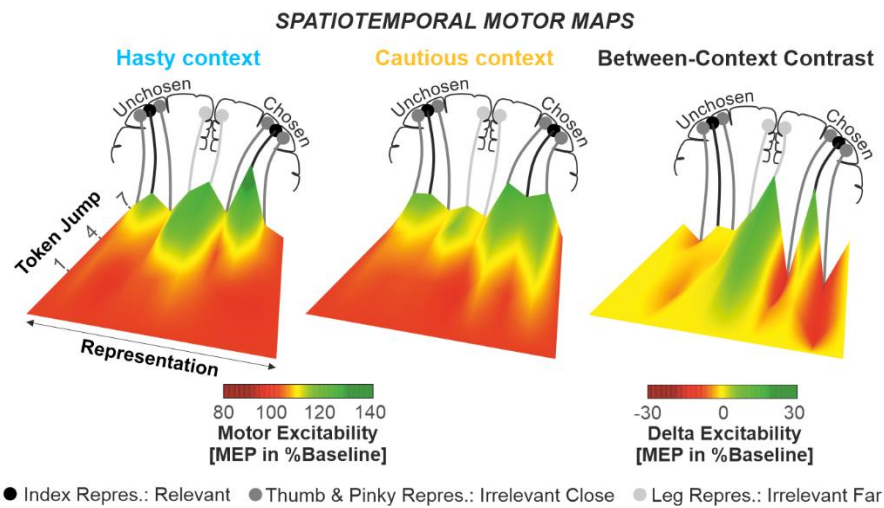

**Figure 5 : La stratégie de décision hâtive est associée à différents types de modulation de l'activité de M1, produisant une amplification large et une suppression de l'excitabilité autour de la représentation exécutant l'action, relativement à la stratégie de prudence** (extrait de Derosiere et al., 2022, *PLoS Biology*). Au sein de ces cartes spatio-temporelles, j'ai organisé les données MEPs en fonction de la somatotopie des représentations du système moteur (axe des abscisses) et des sauts de jetons / du timing de la TMS (axe des ordonnées). La carte de droite représente la différence entre les cartes obtenues dans le contexte hâtif (à gauche) et le contexte de prudence (au centre). On y remarque que la différence d'amplitude des MEPs (en z) est positive pour les représentations de l'index et de la jambe du côté controlatéral à l'effecteur choisi (« chosen side » sur la figure), indiquant une amplification large de l'activité au sein du système moteur. Cependant, cet effet est absent au sein du côté ipsilatéral à l'effecteur choisi (« unchosen side » sur la figure), indiquant que cette amplification n'est pas globale. Par ailleurs, on peut noter que la différence d'amplitude des MEPs est négative pour les représentations des doigts entourant la représentation de l'index, également du côté choisi, ce qui indique une suppression locale de l'excitabilité reflétant potentiellement le recrutement d'un mécanisme d'inhibition latérale.

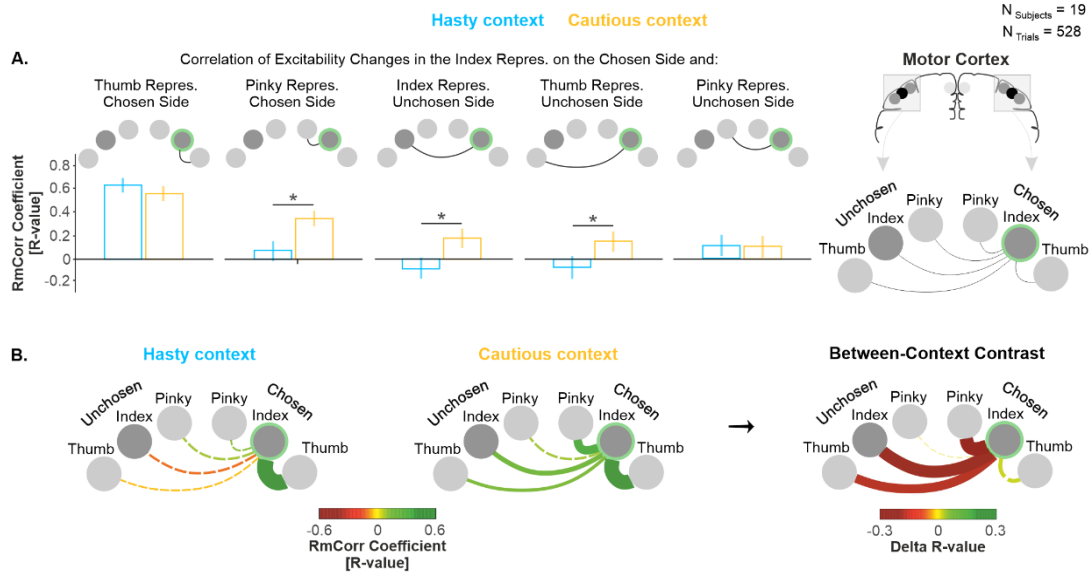

**Figure 6 : La stratégie de décision hâtive est associée à une décorrélation d'activité au sein de M1, relativement à la stratégie de prudence** (extrait de Derosiere et al., 2022, *PLoS Biology*). La carte de corrélation présentée à droite en partie B représente la différence entre les cartes obtenues dans le contexte hâtif (à gauche) et le contexte de prudence (au centre). On y remarque que la différence de valeur de corrélartion (R-value) est négative pour plusieurs pairs de représentations motrices, indiquant une décorrélation des changement d'activité dans le contexte hâtif relativement au contexte de prudence. Ce résultat semble indiquer la présence d'influences affectant la représentation de l'index choisi de manière plus sélective et différenciée dans le contexte hâtif. De manière intéressante, dans le cortex visuel, une décorrélation similaire de l'activité neuronale a été observée lorsque l'attention est dirigée vers un stimulus à l'intérieur du champ réceptif d'une population de neurones. Des analyses computationnelles ont révélé que cette décorrélation due à l'attention améliore considérablement le rapport signal-sur-bruit des signaux neuronaux.

## 2.7. Intégration de signaux de renforcement et de récompense pendant l'apprentissage moteur

Supervisions et publications

Ce projet a impliqué la supervision d'un doctorant et de trois étudiants de Master :

- Pierre Vassiliadis, doctorant co-dirigé avec les Profs. J. Duque et F. Hummel d'octobre 2018 à novembre 2022. Thèse de doctorat sur les mécanismes sous-tendant l'apprentissage moteur renforcé.
- Aegryan Lété, supervisé de juin 2019 à juin 2022. Mémoire de Master sur l'impact du timing de la récompense sur l'apprentissage moteur par renforcement.
- Wanda Materne, supervisée de janvier 2019 à juin 2020. Mémoire de Master sur l'implication de M1 dans l'apprentissage moteur par renforcement.
- Cécile Dubuc, supervisée d'avril 2019 à décembre 2019. Mémoire de Master sur l'impact du contexte de motivation sur l'apprentissage moteur par renforcement.

Articles liés au projet :

- Vassiliadis, Beanato E, Popa T, Windel F, Morishita T, Neufeld E, Duque J, **Derosiere G**, Wessel MJ, Hummel FC. (2022). Non-invasive stimulation of the human striatum disrupts reinforcement learning of motor skills. *Preprint sur BioRxiv* 2022.11.07.515477; doi: https://doi.org/10.1101/2022.11.07.515477 [15].
- Vassiliadis P, Lete A, Duque J, **Derosiere G**. (2022). Reward timing matters in motor learning. *iScience*, 25(5), 104290 [91].
- Vassiliadis P, **Derosiere G**, Lete A, Dubuc C, Crevecoeur F, Hummel F, Duque J. (2021). Reward boosts reinforcement-based motor learning. *iScience*, 24(7), 102821 [92].

- Vassiliadis P, **Derosiere G**. (2020). Selecting and executing actions for rewards. *The Journal of Neuroscience*, 40(34), 6474-6478.7 [28].
- Vassiliadis P, **Derosiere G**, Grandjean J, Duque J. (2020). Motor training strengthens preparatory suppression during movement preparation. *Journal of Neurophysiology*, 124(6), 1656-1666 [93].
- Vassiliadis P, **Derosiere G**, Duque J. (2019). Beyond motor noise: considering other causes of impaired reinforcement learning in cerebellar patients. *eNeuro*, 6(1) [94].

Description du projet

Un étudiant de Master que j'ai co-supervisé (avec la Prof. J. Duque) au début de mon postdoctorat – Pierre Vassiliadis – a développé un intérêt particulier pour la motivation, l'apprentissage par renforcement et l'intégration des signaux de récompense par le système moteur pendant la prise de décision. Pierre a ensuite souhaité poursuivre en doctorat autour de ces thématiques, qu'il a alors appliquées au domaine de l'apprentissage moteur.

L'apprentissage moteur permet aux animaux, y compris aux êtres humains, d'acquérir des compétences essentielles pour interagir efficacement avec l'environnement. Cette capacité à apprendre de nouvelles habiletés motrices est d'une grande importance pratique pour les activités de la vie quotidienne (comme lors de l'apprentissage de la conduite), mais aussi pour la rééducation motrice après une lésion du système nerveux (comme suite à un accident vasculaire cérébral). Pendant longtemps, l'apprentissage moteur a été principalement conceptualisé comme un processus permettant de corriger itérativement des mouvements sur la base d'informations sensorielles (*e.g.*, visuelles ou somatosensorielles). Cependant, au cours des dernières années, il a été reconnu que l'apprentissage moteur est aussi le fruit d'autres mécanismes, dont l'apprentissage par renforcement, un processus par lequel les actions appropriées sont sélectionnées grâce à des informations sur le résultat des mouvements passés (*e.g.*, succès ou échec). À ce titre, des données récentes montrent que le renforcement et la motivation peuvent être bénéfiques pour l'apprentissage moteur, tant chez les personnes en bonne santé que chez les populations neurologiques. Malgré l'importance potentielle de ces résultats pour améliorer les protocoles de revalidation actuels, les mécanismes qui sous-tendent les améliorations liées au renforcement dans l'apprentissage moteur restent largement inexplorés.

Le doctorat de Pierre visait à fournir une compréhension mécanistique plus approfondie de l'apprentissage moteur par renforcement par le biais d'analyses comportementales, de neuroimagerie et de stimulation cérébrale non-invasive. Dans une première étude [92], nous avons découvert que le fait d'augmenter la motivation pendant un entraînement moteur (en offrant une récompense monétaire pour une bonne performance) peut conduire à des améliorations persistantes de la performance qui ne sont pas obtenues avec le feedback de renforcement uniquement, et qui sont liées à une régulation accrue de la variabilité motrice basée sur les résultats des mouvements précédents. Dans une seconde étude, que j'ai supervisée et dont je suis le dernier auteur [95], nous avons étudié l'effet du timing du renforcement (*i.e.*, le délai entre la fin de l'exécution de l'action et le renforcement) sur l'apprentissage moteur et avons découvert que le fait de retarder le renforcement de quelques secondes seulement pouvait fortement influencer la dynamique et la consolidation de l'apprentissage moteur (Figure 7). Enfin, dans une troisième étude [15], nous avons étudié le rôle causal du striatum dans l'apprentissage moteur par renforcement. Ici, nous avons montré, en combinant une nouvelle approche de stimulation cérébrale non-invasive profonde appelée stimulation électrique transcrânienne par interférence temporelle (*i.e.*, Temporal Interference Stimulation ou TIS) et la neuroimagerie, qu'un mécanisme spécifique s'appuyant sur les oscillations haut gamma dans le striatum est impliqué de manière causale dans l'apprentissage moteur par renforcement.

Dans l'ensemble, le travail de doctorat de Pierre Vassiliadis caractérise des mécanismes clés qui sous-tendent l'effet du renforcement sur l'apprentissage moteur, ouvrant la voie à l'incorporation de renforcements optimisés dans les protocoles de rééducation motrice.

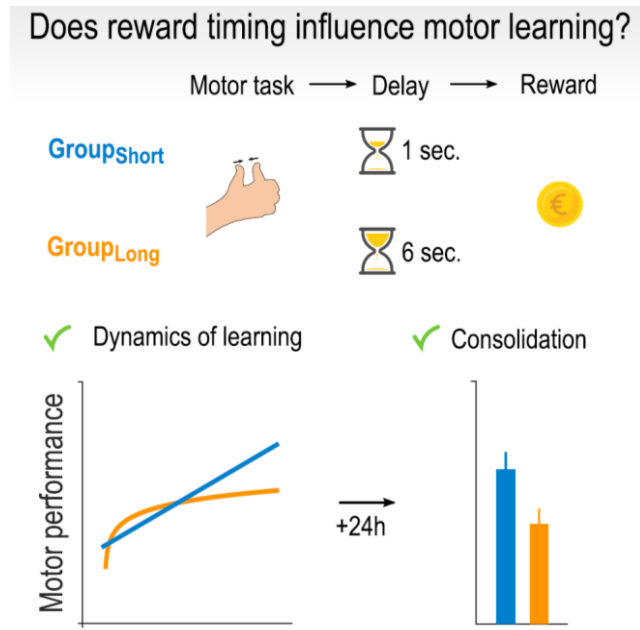

**Figure 7 : Retarder la récompense de quelques secondes après l'exécution de l'action influence la dynamique et la consolidation de l'apprentissage moteur** (extrait de Vassiliadis, *[…]*, Derosiere, 2022, *iScience*)**.** Des travaux sur la prise de décision et l'apprentissage par renforcement a montré que des délais de récompense courts et longs (*e.g.*, 1 et 6 s après l'exécution de l'action) activent de manière préférentielle le striatum et l'hippocampe, respectivement. Étant donné le rôle fonctionnel distinct de ces deux zones dans l'apprentissage moteur, nous avons émis l'hypothèse que le timing de la récompense pourrait moduler la façon dont les gens apprennent et consolident une nouvelle habileté motrice. Nos résultats montrent que le fait de retarder de quelques secondes la récompense influence la dynamique de l'apprentissage moteur. En effet, comme représenté sur le graphique en bas à gauche de la figure, l'entraînement avec un délai de récompense court (1 s ; Group$_{Short}$, en bleu) induit des gains de performance lents mais continus, tandis qu'un délai de récompense long (6 s ; Group$_{Long}$, en orange) conduit à des taux d'apprentissage initialement élevés, suivis d'un plateau précoce dans la courbe d'apprentissage et d'une performance plus faible à la fin de l'entraînement. De plus, le graphique en barres en bas à droite de la figure montre que les sujets ayant bénéficié d'un délai de récompense court montrent une consolidation pendant la nuit, tandis que ceux qui se sont entraînés avec un délai de récompense long ne consolident pas la mémoire motrice. Dans l'ensemble, nos données montrent que le timing de la récompense affecte l'apprentissage moteur, potentiellement en modulant l'engagement de différents processus d'apprentissage, une découverte qui pourrait être exploitée dans de futurs programmes de réhabilitation.

## 2.8. Circuits cérébraux modulant le système moteur durant des processus cognitifs

Supervisions et publications

Ce projet a impliqué la supervision de deux chercheurs visiteurs (un professeur junior et une postdoctorante juniore) et de quatre étudiants de Master :

- Prof. Matthieu Boisgontier, formé aux différents protocoles de TMS de septembre à novembre 2021.
- Dr. Cécilia Neige, postdoctorante supervisée d'avril à juillet 2021 puis de janvier à mars 2022, sur un projet ayant pour objectif de tester l'influence de SMA et de vmPFC sur M1 au repos au travers d'un protocole de ppTMS.
- Thomas Brees, supervisé de janvier 2021 à juin 2022. Mémoire de Master en binôme avec A. Ali Zazou sur le rôle des circuits SMA-M1 et vmPFC-M1 cortico-corticaux et cortico-souscortico-corticaux dans l'apathie chez des sujets sains.
- Abdelkrim Ali Zazou, supervisé de janvier 2021 à juin 2022. Mémoire de Master en binôme avec T. Brees.
- Ahmad Nourredine, supervisé de septembre à décembre 2021 (stage pratique en laboratoire sans mémoire à réaliser).
- Valentin Touzé, supervisé de septembre à décembre 2021 (stage pratique en laboratoire sans mémoire à réaliser).

Articles liés au projet :

- Neige C, Zazou AA, Vassiliadis P, Lebon F, Brees T, **Derosiere G**. (2022). Probing the influence of SMA and vmPFC on the motor system with dual-site transcranial magnetic stimulation. *Preprint sur BioRxiv*: https://doi.org/10.1101/2022.01.18.476729 [96].

- Wilhelm E, Quoilin C, **Derosiere G**, Paco S, Jeanjean A, Duque J. (2022). Corticospinal suppression underlying intact movement preparation fades in late Parkinson's disease. *Movement Disorders* [97].
- **Derosiere G**, Duque J. (2020). Tuning the corticospinal system: how distributed brain circuits shape human actions. *The Neuroscientist*, p. 1073858419896751 [98].

Description du projet

La vision parallèle décrite en section 2.2. postule que les structures sensorimotrices contribuent à différents processus d'ordre cognitif, comme l'intégration de signaux de récompense pendant l'apprentissage et la prise de décision. Je pense que la plupart des scientifiques adoptant cette vision se sont dans un premier temps essentiellement attachés à tester le rôle des structures sensorimotrices dans différents processus cognitifs, isolément des autres structures cérébrales. C'est par exemple mon cas. En effet, comme présenté en sections 2.3, 2.4 et 2.5, la majorité de mon travail de postdoctorat a été focalisée sur l'intégration de variables dites décisionnelles par le système moteur.

Cependant, la vision parallèle ne postule pas que les structures sensorimotrices implémenteraient à elles seules les processus cognitifs mais, au contraire, qu'elles seraient impliquées dans ces processus « en parallèle » d'autres structures fronto-pariétales et sous-corticales, avec lesquelles elles interagiraient continuellement au travers de boucles récurrentes. Ainsi, une approche complémentaire afin de tester la validité de cette vision parallèle consiste à explorer l'influence de diverses structures cérébrales sur les structures motrices *durant les processus cognitifs*, en dehors de toute exécution motrice. En ce sens, au cours des trois dernières années, j'ai pris un certain recul par rapport aux travaux de postdoctorat que j'avais effectués jusqu'alors, et j'ai développé un intérêt pour les circuits cérébraux qui seraient à l'origine des modulations observées au sein du système moteur.

J'ai d'abord effectué un travail de revue de littérature sur ce sujet, que nous avons publié avec Julie Duqué dans *The Neuroscientist* (Derosiere et Duque, 2020). Cet article de revue avait pour objectif de synthétiser les résultats des études ayant exploité la TMS chez l'Homme afin de quantifier l'influence de différents types de circuits sur la voie corticospinale pendant la sélection, l'inhibition et l'arrêt d'actions motrices. En effet, l'utilisation de différentes approches de ppTMS et leur combinaison avec des protocoles de stimulation sous-corticale permet d'investiguer comment des circuits intra-corticaux, trans-corticaux et sous-cortico-corticaux contribuent au contrôle de l'Action en modulant l'activité de la voie corticospinale. Ce travail de revue nous a permis de souligner d'importants manques dans le domaine, qui découlent du fait que ces circuits, et leur impact sur la voie corticospinale, ont fait l'objet d'un nombre restreint d'études et n'ont donc pas été considérés de manière équivalente pour les processus de sélection, d'inhibition et d'arrêt des actions motrices. Cela a conduit à l'idée trompeuse que certains circuits ou régions sont spécialisés dans des processus spécifiques et qu'ils produisent des modulations particulières de l'excitabilité corticospinale (*e.g.*, modulation globale ou spécifique de l'excitabilité corticospinale ; Figure 8). Par conséquent, nous soulignons dans cet article de revue la nécessité d'adopter des approches de recherche plus transversales dans le domaine.

Sur le plan expérimental, j'ai dirigé une étude qui avait pour objectif d'établir l'applicabilité de la ppTMS à double-site sur certaines aires du cortex frontal médian dont l'aire motrice supplémentaire (SMA) et le vmPFC ; je suis le dernier auteur de l'article résultant de cette étude, actuellement pré-publié sur *BioRxiv* [96]. Cette étude est liée au projet de recherche que je présente en section 3, dont l'un des objectifs est d'exploiter la ppTMS à double-site afin de quantifier le rôle de circuits connectant la SMA et le vmPFC à M1 dans la prise de décision. Cependant, plusieurs questions restaient ouvertes quant à l'utilisation de la ppTMS à double-site sur ces aires. Tout d'abord, les quelques études TMS qui avaient ciblé la SMA jusque maintenant s'étaient surtout focalisées sur des intervalles inter-stimulation courts (6-8 ms), supposés recruter les circuits cortico-corticaux [100,101]. De ce fait, on manquait de données sur la nature de l'influence (*i.e.*, facilitatrice ou suppressive) de la stimulation de la SMA sur M1 lorsqu'elle est appliquée avec des intervalles plus longs (10-15 ms), supposés recruter des circuits cortico-sous-corticaux plus indirects. Deuxièmement, il n'était pas clair si l'influence facilitatrice de la stimulation de la SMA rapportée précédemment avec des intervalles courts reflète le recrutement de circuits cortico-corticaux (comme on le suppose généralement) ou résulte de la sommation, au niveau spinal, de volées descendantes au sein des cellules pyramidales provenant de la SMA et de M1. Troisièmement, la TMS à double-site n'avait jamais été utilisée jusqu'à présent pour sonder l'influence du vmPFC sur M1, probablement en raison de la difficulté présumée d'atteindre le vmPFC avec le champ magnétique. Nos résultats montrent que la stimulation de la SMA facilite l'activité de M1 avec un intervalle long de 12 ms. De plus, nos données révèlent que l'influence facilitatrice de la stimulation de SMA observée avec des intervalles courts ne résulte pas d'interactions spinales. Enfin, nous montrons que la stimulation du vmPFC induit un effet suppressif modéré sur M1, que ce soit avec des intervalles inter-stimulation courts ou longs. Dans l'ensemble, cette étude nous a permis de montrer la faisabilité de quantifier l'activité de circuits SMA-M1 et vmPFC-M1 cortico-corticaux et cortico-sous-cortico-corticaux, une approche qui sera exploitée pendant la prise de décision au sein du projet présenté en section 3.

Enfin, je suis également directement impliqué dans deux autres études, menées par la Prof. J. Duque à Bruxelles, dont l'objectif est de déterminer le rôle des ganglions de la base dans les modulations d'activité de M1 pendant l'inhibition de l'action. Les résultats de la première étude ont été publiés au sein de la revue *Movement Disorders* et suggèrent que l'accroissement de la neurodégénérescence des ganglions de la base avec l'avancée de la maladie de Parkinson réduit la suppression de l'activité de M1 généralement observée pendant les phases d'inhibition de l'action [97]. De manière intéressante, nos données montrent que cette réduction de suppression de M1 est associée à l'aggravation de la bradykinésie, un des symptômes cardinaux de cette pathologie. La seconde étude a pour objectif d'explorer le rôle causal du noyau sous-thalamique dans les modulations d'activité de M1 pendant l'inhibition motrice. L'analyse des résultats de cette seconde étude est toujours en cours et la soumission de l'article, dont je suis le deuxième auteur, est prévue pour mai 2023.

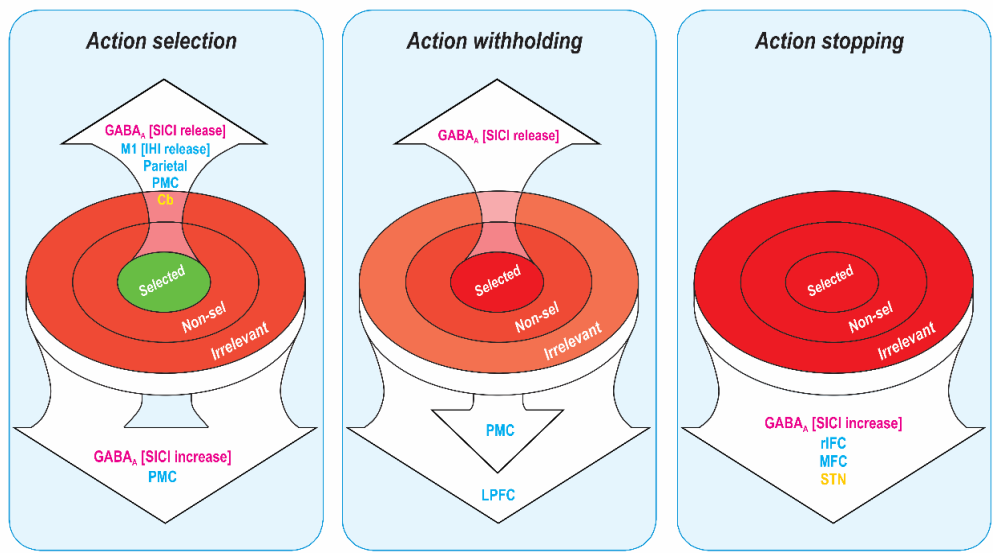

**Figure 8 : Synthèse graphique des changements d'excitabilité corticospinale survenant au cours de la sélection (gauche), l'inhibition (centre) et l'arrêt (droite) d'actions motrices** (extrait de Derosiere et Duque, 2020, *The Neuroscientist*). Les cercles représentent les représentations motrices sélectionnées, non-sélectionnées et non-pertinentes pour la tâche, tandis que leur couleur représente les changements nets de l'excitabilité corticospinale (*i.e.*, rouge = suppression, vert = facilitation). Comme évident sur cette figure, chacun des circuits, et leur impact sur la voie corticospinale, ont fait l'objet d'un nombre restreint d'études TMS et n'ont donc pas été considérés de manière équivalente pour les processus de sélection, d'inhibition et d'arrêt des actions motrices. Par exemple, les circuits transcorticaux pariéto-moteurs ont été seulement étudiés dans le contexte de sélection motrice mais pas dans les contextes d'inhibitions et d'arrêt d'actions. Qui plus est, dans le contexte de sélection motrice, l'influence de ce circuit sur la voie corticospinale, n'a pas été étudiée pour les représentations motrices non-sélectionnées et non-pertinentes pour la tâche. Cela a conduit à l'idée trompeuse que certains circuits ou régions sont spécialisés dans des processus spécifiques et qu'ils produisent des modulations particulières de l'excitabilité corticospinale (*e.g.*, modulation globale ou spécifique de l'excitabilité corticospinale). Par conséquent, nous soulignons dans cet article de revue la nécessité d'adopter des approches de recherche plus transversales dans le domaine.

## 2.9. Recherches passées : conclusion intermédiaire

En résumé, l'un des principaux objectifs de mes recherches jusqu'à ce jour a été de mieux comprendre la contribution du système sensorimoteur dans les processus cognitifs. La Figure 9, ci-dessous, synthétise de manière schématique les projets que j'ai menés jusque maintenant.

Comme décrit dans mon projet de recherche ci-dessous, je souhaite maintenant poursuivre ma propre ligne de recherche autour de cette question, avec pour objectif principal de caractériser le rôle fonctionnel des circuits fronto-striato-moteurs dans la prise de décision basée sur l'effort et l'apathie comportementale.

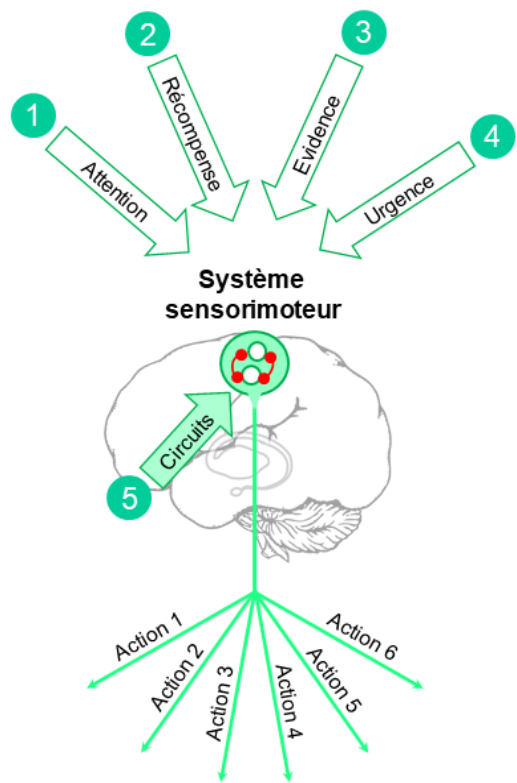

**Figure 9 : Synthèse du cadre théorique et des cinq projets principaux que j'ai menés au cours des douze dernières années.**

# 3. Projet de recherche

Des insectes aux humains en passant par les rongeurs, les animaux doivent constamment décider s'il vaut la peine de s'engager (ou non) dans des efforts physiques afin d'atteindre des récompenses présentes dans l'environnement. Ce processus – généralement appelé prise de décision basée sur l'effort – est dysfonctionnel dans l'apathie, un syndrome dont la composante comportementale est caractérisée par une propension réduite à entreprendre des actions requérant des efforts, et dont la prévalence est élevée dans divers troubles neurologiques et psychiatriques, comme la maladie de Parkinson (40 %), la maladie d'Alzheimer (49 %), les accidents vasculaires cérébraux (39 %), les démences frontotemporale et vasculaire (72 et 65 %), la schizophrénie (47 %) et la dépression (38 %) [102,103]. Aujourd'hui, d'importantes recherches sont nécessaires afin de déterminer les causes neuronales de l'apathie comportementale. Un corollaire direct à cela est la nécessité d'identifier les réseaux cérébraux et les mécanismes neurophysiologiques qui sous-tendent les décisions basées sur l'effort. Ce type de recherche fournira une base de connaissance qui permettra à terme de développer des interventions visant à réduire l'apathie comportementale chez les patients (Husain and Roiser, 2018), par exemple au travers d'approches de stimulation cérébrale.

Les études antérieures sur le sujet ont abouti à trois grandes conclusions. D'abord, la prise de décision basée sur l'effort est généralement associée à une augmentation d'activité au sein de la SMA, du vmPFC et du striatum [104–109]. Ensuite, chacune de ces structures semble être impliquée dans des processus spécifiques de la prise de décision. En effet, alors que l'activité de la SMA est essentiellement corrélée au niveau d'effort attendu (*e.g.*, Croxson et al., 2009), l'activité du vmPFC covarie préférentiellement avec la magnitude de la récompense (Husain and Roiser, 2018 ; Kroemer et al., 2016 ; Pessiglione et al., 2018, voir cependant : Hogan et al., 2019), suggérant que ces structures évaluent l'effort à réaliser et la récompense en jeu, respectivement (Kroemer et al., 2016). Le striatum, quant à lui, présente des changements d'activité dépendants à la fois de l'effort et de la récompense, et est donc souvent considéré comme un centre intégratif, évaluant le rapport entre les coûts et les bénéfices des choix d'action (Kroemer et al., 2016 ; Suzuki et al., 2020). Enfin, l'atrophie et/ou la dysfonction de ces structures peut conduire à l'apathie [102,110–112], ce qui suggère que ces dernières joueraient un rôle causal dans la prise de décision basée sur l'effort.

La vision prépondérante dans ce champ d'étude est que les décisions basées sur l'effort émergent d'interactions en boucle fermée au sein de ce réseau fronto-striatal. En effet, comme mentionné en section 2.2., les bases neurales de la prise de décision ont été historiquement considérées au travers d'un cadre théorique considérant le comportement comme résultant d'une progression sérielle de processus perceptif, cognitif et moteur [113]. Dans cette optique, le réseau fronto-striatal décrit ci-dessus implémenterait à lui seul le processus cognitif – la prise de décision basée sur l'effort – avant de signaler aux structures du système moteur, comme M1, l'ordre d'exécuter l'action si nécessaire. Cependant, certaines données expérimentales ont remis en question la validité de cette vision sérielle. Par exemple, certains des projets que j'ai décrits en section 2 ont révélé que l'activité motrice évoquée (Derosiere et al., 2018, *The Journal of Neuroscience*), l'activité motrice oscillatoire (Alamia, [...], Derosiere, 2019, *NeuroImage*) et l'excitabilité corticospinale (Derosiere et al., 2022, *PLOS Biology*), toutes enregistrées au sein de M1, varient en fonction de variables dites décisionnelles pendant la délibération (*e.g.*, en fonction du niveau d'évidence sensorielle en faveur du choix), alors même qu'il n'y a aucun mouvement à exécuter. Comme décrit plus haut, ce type de résultat a conduit à proposer une "vision parallèle", selon laquelle les changements d'activité observés au sein du système moteur joueraient un rôle décisif dans le processus de décision et détermineraient si une action est éventuellement sélectionnée et exécutée ou non. Une variété de variables biaiserait ces changements d'activité motrice, rapprochant ou éloignant l'activité du seuil de déclenchement, et augmentant ou diminuant donc la probabilité d'initier certaines actions. Ces variables incluent notamment l'effort attendu et la récompense en jeu.

Conformément à cette idée, diverses études menées chez les insectes [114], les rongeurs [115] et l'Homme [38,39] ont maintenant montré que les changements d'activité motrice enregistrés pendant la prise de décision varient en fonction des niveaux d'effort et de récompense attendus. Par exemple, certains chercheurs, qui ont appliqué la spTMS sur M1 pour mesurer l'excitabilité corticospinale, ont observé que l'amplitude des MEPs augmente plus rapidement avant d'initier des actions associées à de faibles coûts biomécaniques, nécessitant peu d'effort (*i.e.*, par rapport à des actions impliquant un coût élevé) [116]. Similairement, l'activité motrice augmente plus rapidement lorsque les sujets choisissent des actions associées à des récompenses élevées (*i.e.*, par rapport à des actions impliquant une récompense faible) [38,39]. De manière surprenante, l'origine de ces modulations reste inconnue à l'heure actuelle. Cependant, des patterns d'activité très similaires se produisent au sein de la SMA, du vmPFC et du striatum, également pendant la phase de décision. En effet, la SMA présente une activité accrue pour des niveaux décroissants d'effort attendu, le vmPFC une activité accrue pour des récompenses plus importantes, et le striatum présente les deux effets, ce qui suggère qu'un lien fonctionnel entre ces structures et M1 pourrait constituer un mécanisme central de la prise de décision basée sur l'effort.

De récents progrès dans le domaine de la stimulation cérébrale pourraient permettre de tester cette hypothèse novatrice directement chez l'Homme. En effet, alors que la ppTMS à double-site permet d'évaluer l'influence causale des zones frontales sur M1 [96], la stimulation par interférence temporelle (TIS) [15,117–119], une technique récente, offre la possibilité de cibler des structures cérébrales profondes, comme le striatum, de manière non-invasive, et donc d'étudier leur rôle fonctionnel dans des réseaux cérébraux spécifiques. La TIS consiste à délivrer deux champs électriques à la surface du crâne par le biais de deux paires d'électrodes à des fréquences trop élevées pour recruter les neurones au niveau cortical (*e.g.*, 2 kHz), mais qui diffèrent l'une de l'autre d'un delta compris dans le spectre naturel de décharge des neurones (*e.g.*, 50 Hz). La superposition des deux champs électriques dans la boîte crânienne donne lieu à un champ électrique cumulé, dont l'enveloppe est modulée en amplitude à cette valeur de delta de fréquence. La modulation de l'enveloppe à un endroit particulier dépend de la somme vectorielle des deux champs appliqués à ce point et, par conséquent, peut avoir un maximum à un point éloigné des électrodes, au niveau des tissus profonds du cerveau. Comme indiqué en section 2.6, j'ai exploité cette technique avec certains de mes collaborateurs dans le cadre d'une cotutelle de thèse (*i.e.*, avec le Prof. F. Hummel à l'EPFL, Genève) et la modélisation du champ électrique ainsi que des données d'imagerie par résonance magnétique fonctionnelle (IRMf) obtenues chez 14 sujets démontrent l'efficacité de cette technique pour moduler l'activité striatale (voir la Figure 12.E). La résolution spatiale actuelle de la TIS permet de recruter simultanément les parties dorsale et ventrale du striatum, où se projettent respectivement la SMA et le vmPFC.

L'objectif de mon projet est de **questionner le rôle des boucles récurrente présentes entre le réseau fronto-striatal décrit ci-dessus et le cortex moteur dans la prise de décision basée sur l'effort et dans l'apathie comportementale**. Au sein de cet objectif large, je testerai dans un premier temps l'hypothèse que la SMA et le vmPFC modulent de manières effort- et récompense-dépendantes l'activité de M1, respectivement, et que leur influence modulatrice implique des circuits convergeant via le striatum. En d'autres termes, je propose que l'activité motrice est modulée par le striatum, qui intègre les signaux effort- et récompense-dépendants en provenance de la SMA et du vmPFC. Ce projet implique une combinaison unique d'approches de neurostimulation: TMS à impulsion simple et double – largement exploitées pour mesurer l'activité de M1 et la connectivité effective chez l'Homme – et TIS, une technique innovante. Il répondra aux trois objectifs suivants : 1. Etablir le rôle causal du striatum dans les modulations effort- et récompense-dépendantes de l'activité de M1, en combinant la TMS à impulsion simple avec la TIS ; 2. Déterminer le rôle de la SMA et du vmPFC dans ces modulations de M1 et leur dépendance vis-à-vis du striatum, en couplant la TMS à impulsion double avec la TIS ; et 3. Mesurer la connectivité effective entre la SMA / le vmPFC et M1 chez les patients apathiques, en utilisant la TMS à impulsion double. Alors que la plupart des recherches passées se sont concentrées exclusivement sur les structures fronto-striatales ou motrices, l'hypothèse holistique centrale de mon projet de recherche appelle à élargir le champ d'investigation aux circuits fronto-striato-moteurs (voir Figure 10, ci-dessous) et implique qu'un dysfonctionnement de ces circuits peut contribuer à l'apathie comportementale. Comme présenté en section *3.4 Perspectives* (ci-après), un objectif de carrière à long terme est de développer des interventions de stimulation cérébrale permettant de moduler la connectivité effective au sein des circuits étudiés, avec pour but ultime de réduire l'apathie comportementale chez les patients.

# 4. Réflexion personnelle

En tant que scientifiques, le propre de notre travail est de produire de la connaissance et de la transmettre. Pour cela, nous choisissons de tester une certaine vision théorique (*e.g.*, la vision parallèle) au travers de grandes questions de travail (*e.g.*, M1 intègre-t-il des variables décisionnelles ?), en exploitant certaines approches paradigmatiques (*e.g.*, en mesurant l'influence de variables multiples sur l'activité de M1). Au cours des dix dernières années, ma perception des questions clés à aborder afin de tester la vision parallèle – et, de ce fait, des approches à exploiter – a progressivement évolué. Ici, je propose une réflexion personnelle sur cette évolution.

### 4.1. De la focalisation sur le système moteur à l'approche en réseau

Une large partie de mes travaux (2014 – 2022) a d'abord questionné le rôle fonctionnel du système moteur dans l'intégration de variables décisionnelles (voir sections 2.3, 2.4 et 2.5). Pour cela, bien que la méthodologie exacte ait pu varier (rTMS, EEG, spTMS, *etc.*), mon approche a consisté à étudier systématiquement soit l'impact d'une perturbation de M1 sur l'intégration de variables dans le comportement décisionnel, soit l'influence de ces variables sur les changements d'activité de M1. Plus récemment (2020 – présent), je me suis questionné sur l'origine neurale des variations d'activité observées au sein de M1 pendant la décision (voir sections 2.7 et 3). Pour répondre à cette question, je propose d'exploiter une approche de « *perturbation-et-mesure* » de l'activité neurale : l'activité d'une structure à distance de M1 est perturbée et l'on mesure l'impact sur les variations d'activité de M1 au cours de la prise de décision. Dans la continuité de cette réflexion, dernièrement (2022 – présent), j'ai remarqué qu'alors qu'un des postulats de base de la vision parallèle est que les structures sensorimotrices interagissent avec d'autres structures de manière *bi-directionnelle*, au travers de circuits récurrents, le rôle fonctionnel de M1 dans les variations d'activité d'autres structures n'a encore jamais été testé. Cela constitue une de mes perspectives de travail principales (voir section 3.4). Ici, une approche possible consisterait à augmenter ou diminuer la connectivité au sein des circuits présents entre M1 et certaines structures cibles (*e.g.*, la SMA ou le vmPFC) en exploitant la ccPAS (voir section 3.4) et à étudier l'impact sur l'activité neurale au sein du réseau considéré et sur le comportement.

Pourquoi avoir, dans un premier temps, choisi de consacrer plusieurs années de recherche à étudier le rôle fonctionnel du système moteur dans l'intégration de variables décisionnelles ? Nous n'étions pas, avec Julie Duqué, les seuls à adopter cette approche, mais nous inscrivions bien dans un courant plus large dans lequel baignaient plusieurs équipes dans le monde. Je pense qu'une partie de ce choix est liée au fait que la vision parallèle de la décision reste encore à un stade relativement précoce de son existence, celle-ci ayant émergé au cours des vingt dernières années. Ludwik Fleck et Thomas Kühn, deux épistémologistes constructivistes du XX ème siècle, ont décrit comment l'émergence d'une nouvelle vision théorique en Science résultait d'un processus cumulatif d'évidence. Alors qu'une vision théorique est prédominante (*e.g.*, la vision sérielle), certains résultats commencent à aller à son encontre, menant à proposer une vision alternative (*e.g.*, la vision parallèle). S'opère alors un shift théorique, qui nous amène progressivement à étudier le monde au travers de nouvelles questions de recherche, en exploitant de nouvelles approches paradigmatiques, ce qui entraîne généralement un cumul d'évidence en défaveur de la vision passée et en faveur de la vision alternative.

Ce que je crois, c'est que les questions abordées et les approches exploitées pendant cette phase instable de shift théorique sont relativement drastiques. L'objectif est alors d'apporter le niveau d'évidence le plus probant en faveur d'une des deux visions en compétition. C'est, je pense, ce qui s'est passé dans notre champ, nous menant à étudier essentiellement l'implication de structures sensorimotrices dans l'intégration de variables multiples pendant la décision. Nous allions ainsi complètement à contre-courant de la vision sérielle. Je pense que cet effet de nouveauté de la vision alternative sur l'approche exploitée est d'autant plus fort que la vision historiquement prépondérante est très ancrée dans le champ. C'était le cas de la vision sérielle, dont l'héritage remonte *a minima* aux psychologues comportementalistes du début du XX ème siècle. Ajouter à cela un système de publication, de citation, et, *in fine,* d'embauche, qui valorise le scoop scientifique, et la recherche d'approches dites « sexy », originales et nouvelles, comme nous en parlons généralement entre nous, devient alors un de nos objectifs principaux. Comme décrit plus haut, je pense qu'il est maintenant temps de se détacher de cette focalisation sur les structures sensorimotrices afin d'identifier comment ces dernières interagissent avec d'autres structures de manière bi-directionnelle, au travers de circuits récurrents, pendant la décision. À mes yeux, répondre à question est tout aussi excitant !

## 4.2. De la pensée en « blanc ou noir » aux 50 nuances de gris

J'ai été plusieurs fois frappé ces dernières années par la tendance que nous avons, en tant que scientifiques, à dichotomiser notre pensée, de telle façon que le nombre d'hypothèses, de modèles et de théories proposés, est bien souvent binaire. Voici certains des exemples que j'ai relevés au cours de mes années postdoctorales : top-down versus bottom-up dans le champ de l'attention, automatique versus contrôlé dans le champ du contrôle cognitif, model-based versus model-free dans le champ de l'apprentissage par renforcement, feedback versus feedforward dans le champ du contrôle moteur, competition resolution versus impulse control dans le champ de la préparation de l'action, exploration versus exploitation dans le champ du foraging, liking versus wanting dans le champ de la motivation, *etc.* En réalité, cette tendance à la pensée dichotomique n'est pas propre aux scientifiques, mais semble être une propriété de la cognition humaine [120–122].

Pourquoi ce constat ? Je crois que la phase de shift théorique mentionnée plus haut exacerbe cette pensée en « blanc ou noir », notamment pour les raisons que j'ai déjà décrites. Cependant, je pense qu'il est maintenant nécessaire d'élargir le champ des propositions et de sortir de cette pensée dichotomique, tout en tempérant ce que l'on entend (ou tout du moins ce que j'entends personnellement) par « vision parallèle de la décision ». D'abord, une certaine temporalité existe indéniablement selon moi dans les processus cognitifs : bien que certains processus se déroulent en parallèle ou se chevauchent fortement dans le temps (*e.g.*, les processus de valuation de la récompense et de préparation du mouvement), certains sont plus enclins à prendre place en phase précoce qu'en phase tardive de la décision. Dans le même sens, la contribution respective de chacune des structures d'un réseau donné varie très probablement au cours du processus de délibération. Ainsi, je pense que, pour évoluer, notre champ ne devrait plus se focaliser sur la confrontation entre la vision sérielle et parallèle (comme c'est encore parfois le cas ; [123–125]), mais plutôt étudier comment le rôle respectif de chacune des structures d'un réseau donné évolue dynamiquement au cours de la décision. En cela, une vision que j'appelle « *parallèlo-sérielle de la décision* » serait peut-être plus adéquate. Ensuite, il faut bien noter que la vision parallèle n'implique pas que toutes les structures d'un réseau donné contribuent avec le même poids dans l'ensemble des processus étudiés. Différentes données neuroanatomiques et neurophysiologiques montrent qu'il existe bel et bien une spécialisation des structures cérébrales ; il est important selon moi de prendre en considération cette spécialisation au sein de notre vision parallèle. Enfin, la contribution des différentes structures d'un réseau varie aussi très probablement en fonction du type de décision à prendre. En effet, bien que toute décision s'exprime finalement par une action, la relation entre décision et action est parfois très indirecte. Par exemple, lorsque l'on décide d'acheter une maison, on ne choisit vraisemblablement pas entre les mouvements pour ouvrir la porte, mais on choisit plutôt entre des quantités abstraites telles que la localisation et le trajet pour se rendre au travail. Dans ce cas, la décision est abstraite par nature et la contribution des structures sensorimotrices est probablement minime. En revanche, lors d'un match de tennis, l'évidence sensorielle fournie par la trajectoire de la balle en mouvement offre aux joueurs de multiples choix d'action. Dans ce scénario, la contribution des structures sensorimotrices impliquées dans le contrôle des effecteurs en jeu est probablement plus importante, leur permettant d'exécuter l'action en une fraction de seconde lorsque cela est nécessaire. Toute situation se situant entre ces scénarios extrêmes devrait impliquer les structures sensorimotrices ainsi que les centres exécutifs plus abstraits avec différents poids, par le biais d'une architecture parallèle. Ainsi, je pense qu'un autre objectif auquel notre champ devra s'atteler dans le futur est de comparer la contribution de différentes structures dans différents types de prise de décision, afin de tester la validité de cette idée.

# 5. Références bibliographiques

# Résumé

Mes travaux de recherche sont ancrés sur le plan théorique dans une vision parallèle de la Cognition, qui postule que les régions sensorimotrices sont impliquées – en parallèle à d'autres structures – dans des processus cognitifs historiquement principalement associés à des régions cérébrales d'ordre supérieur (*e.g.*, processus décisionnels associés au cortex préfrontal). Sur le plan méthodologique, les projets que j'ai menés ont impliqué l'utilisation de plusieurs techniques chez l'Homme (*e.g.*, la stimulation magnétique transcrânienne à impulsion unique, pairée et répétitive, l'électromyographie, l'électroencéphalographie, la spectroscopie dans le proche infrarouge, l'Imagerie par Résonance Magnétique, la stimulation par interférence temporelle) ainsi que diverses approches d'analyse des données (*e.g.*, analyses spatiotemporelles des potentiels évoqués moteurs, de permutations de Monte-Carlo sur des données d'électroencéphalographie, d'apprentissage machine sur des données de neuroimagerie).

Une large partie de mes travaux (2014 – 2022) a d'abord questionné le rôle fonctionnel du système moteur dans l'intégration de variables dites décisionnelles telles que la récompense associée aux différentes actions, l'évidence sensorielle en faveur de chacune des actions, et le niveau d'urgence dans un contexte donné. Pour cela, bien que la méthodologie exacte ait pu varier, mon approche a consisté à étudier systématiquement soit l'impact d'une perturbation du cortex moteur primaire (M1) sur l'intégration de variables dans le comportement décisionnel, soit l'influence de ces variables sur les changements d'activité de M1 pendant la décision. Plus récemment (2020 – présent), je me suis questionné sur l'origine neurale des variations d'activité observées au sein de M1 pendant la décision. Pour répondre à cette question, je propose d'exploiter une approche de « *perturbation-et-mesure* » : l'activité d'une structure à distance de M1 est perturbée et l'on mesure l'impact sur les variations d'activité de M1 au cours de la prise de décision. Dans la continuité de cette réflexion, dernièrement (2022 – présent), j'ai remarqué qu'alors qu'un des postulats de base de la vision parallèle est que les structures sensorimotrices interagissent avec d'autres structures de manière bi-directionnelle, au travers de circuits récurrents, le rôle fonctionnel de M1 dans les variations d'activité d'autres structures n'a encore jamais été testé. Cela constitue une de mes perspectives de travail principales. Je propose une réflexion personnelle concernant cette évolution paradigmatique et discute ma vision des grandes questions à aborder pour notre champ de recherche.